\documentclass{article}

\usepackage{graphicx,hyperref}
\usepackage{amsthm,amsfonts,amsopn,amsmath,amssymb,natbib}
\usepackage[latin1]{inputenc}
\usepackage{xspace}
\usepackage{float}
\usepackage{algorithm,algorithmic,color}
\usepackage[section]{placeins}
\usepackage{sidecap}
\usepackage[top=1.93in, bottom=1.6in, left=1.6in, right=1.6in]{geometry}
\usepackage{microtype}

\DeclareMathOperator{\Var}{Var}

\DeclareMathOperator{\Cov}{Cov}
\DeclareMathOperator{\Id}{I}
\newcommand{\npar}{\par \vspace{2.3ex plus 0.3ex minus 0.3ex}}
\newcommand{\E}{\mathbb{E}}

\newcommand{\var}{\mathrm{var}}
\theoremstyle{plain}

\newcommand{\cites}[1]{\citeauthor{#1}'s (\citeyear{#1})}

\def\var{\mathop{\rm Var}} 
\def\cov{\mathop{\rm Cov}} 
\def\E{\mathbb{E}} 

\author{Tim Salimans\footnote{Erasmus University Rotterdam
 \href{mailto:salimans@ese.eur.nl}{salimans@ese.eur.nl}}\ ~and David A. Knowles\footnote{Stanford University
 \href{mailto:dak33@stanford.edu}{dak33@stanford.edu}}}
 
\title{Fixed-Form Variational Posterior Approximation through Stochastic Linear Regression}

\begin{document}

\renewcommand{\thefootnote}{\fnsymbol{footnote}}

\maketitle

\renewcommand{\thefootnote}{\arabic{footnote}}

\begin{abstract} We propose a general algorithm for approximating nonstandard Bayesian posterior distributions. The algorithm minimizes the Kullback-Leibler divergence of an approximating distribution to the intractable posterior distribution. Our method can be used to approximate any posterior distribution, provided that it is given in closed form up to the proportionality constant. The approximation can be any distribution in the exponential family or any mixture of such distributions, which means that it can be made arbitrarily precise. Several examples illustrate the speed and accuracy of our approximation method in practice.
\end{abstract}

\section{Introduction}
\label{intro}
In Bayesian analysis the form of the posterior distribution is often not analytically tractable. To obtain quantities of interest under such a distribution, such as moments or marginal distributions, we typically need to use Monte Carlo methods or approximate the posterior with a more convenient distribution. A popular method of obtaining such an approximation is \textit{structured} or \textit{fixed-form Variational Bayes}, which works by numerically minimizing the Kullback-Leibler divergence of an approximating distribution in the exponential family to the intractable target distribution~\citep{Attias2000,Beal2006,jordan1999introduction,wainwright2008graphical}. For certain problems, algorithms exist that can solve this optimization problem in much less time than it would take to approximate the posterior using Monte Carlo methods \citep[see e.g.][]{honkela}. However, these methods usually rely on analytic solutions to certain integrals and need conditional conjugacy in the model specification, i.e. the distribution of each variable conditional on its Markov blanket must be an analytically tractable member of the exponential family for these methods to be applicable. As a result this class of methods is limited in the type of approximations and posteriors they can handle.
\npar
We show that solving the optimization problem of fixed-form Variational Bayes is equivalent to performing a linear regression with the sufficient statistics of the approximation as explanatory variables and the (unnormalized) log posterior density as the dependent variable. Inspired by this result, we present an efficient stochastic approximation algorithm for solving this optimization problem. In contrast to earlier work, our approach does not require any analytic calculation of integrals, which allows us to extend the fixed-form Variational Bayes approach to problems where it was previously not applicable. Our method can be used to approximate any posterior distribution, provided that it is given in closed form up to the proportionality constant. The type of approximating distribution can be any distribution in the exponential family or any mixture of such distributions, which means that our approximations can in principle be made arbitrarily precise. While our method somewhat resembles performing stochastic gradient descent on the variational objective function in parameter space~\citep{paisleyvb,nottstochastic}, the linear regression view gives insights which allow a more computationally efficient approach.
\npar
Section~\ref{vb} introduces fixed-form variational posterior approximation, the optimization problem to be solved, and the notation used in the remainder of the paper. In Section~\ref{linreg} we provide a new way of looking at variational posterior approximation by re-interpreting the underlying optimization as a linear regression problem. We propose a stochastic approximation algorithm to perform the optimization in Section~\ref{sa}. In Section~\ref{marglike} we discuss how to assess the quality of our posterior approximations and how to use the proposed methods to approximate the marginal likelihood of a model. These sections represent the core ideas of the paper.
\npar
To make our approach more generally applicable and computationally efficient we provide a number of extensions in two separate sections. Section~\ref{simul} discusses modifications of our stochastic approximation algorithm to improve efficiency. Section~\ref{mix} relaxes the assumption that our posterior approximation is in the exponential family, allowing instead mixtures of exponential family distributions. Sections \ref{sa}, \ref{simul}, and \ref{mix} also contain multiple examples of using our method in practice, and show that despite its generality, the efficiency of our algorithm is highly competitive with more specialized approaches. Code for these examples is available at \url{github.com/TimSalimans/LinRegVB}. Finally, Section~\ref{conclusion} concludes.

\section{Fixed-form Variational Bayes}
\label{vb}

Let $x$ be a vector of unknown parameters and/or latent random effects for which we have specified a prior distribution $p(x)$, and let $p(y|x)$ be the likelihood of observing a given set of data, $y$. Upon observing $y$ we can use Bayes' rule to obtain our updated state of belief, the posterior distribution
\begin{equation}
\label{eq:post1}
p(x|y) = \frac{p(x,y)}{p(y)} = \frac{p(y|x)p(x)}{\int p(y|x)p(x) dx}.
\end{equation}
An equivalent definition of the posterior distribution is
\begin{equation}
\label{eq:post2}
p(x|y) = \arg\min_{q(x)} \E_{q(x)}\left[\log \frac{q(x)}{p(x,y)}\right] = \arg\min_{q(x)}D[q(x)|p(x|y)], 
\end{equation}
where the optimization is over all proper probability distributions $q(x)$, and where \\* $D[q(x)|p(x|y)]$ denotes the Kullback-Leibler divergence between $q(x)$ and $p(x|y)$. The KL-divergence is always non-negative and has a unique minimizing solution $q(x) = p(x|y)$ almost everywhere, at which point the KL-divergence is zero. The solution of (\ref{eq:post2}) does not depend on the normalizing constant $p(y)$ of the posterior distribution.
\npar
The posterior distribution given in (\ref{eq:post1}) is the exact solution of the \textit{variational} optimization problem in (\ref{eq:post2}), but except for certain special cases it is not very useful by itself because it does not have an analytically tractable form. This means that we do not have analytic expressions for the posterior moments of $x$, the marginals $p(x_{i}|y)$, or the normalizing constant $p(y)$. One method of solving this problem is to approximate these quantities using Monte Carlo simulation. A different approach is to restrict the optimization problem in (\ref{eq:post2}) to a reduced set of more convenient distributions $Q$. If $p(x,y)$ is of conjugate exponential form, choosing $Q$ to be the set of factorized distributions $q(x)=q(x_{1})q(x_{2}) \dots q(x_{k})$ often leads to a tractable optimization problem that can be solved efficiently using an algorithm called Variational Bayes Expectation Maximization \citep[VBEM,][]{beal2003vbem}. Such a factorized solution is attractive because it makes the variational optimization problem easy to solve, but it is also very restrictive: it requires a conjugate exponential model and prior specification and it assumes posterior independence between the different blocks of parameters $x_{i}$. This means that this factorized approach can be used with few models, and that the solution $q(x)$ may be a poor approximation to the exact posterior \citep[see e.g.][]{Turner}.
\npar
An alternative choice for $Q$ is the set of distributions of a certain parametric form $q_{\eta}(x)$, where $\eta$ denotes the vector of parameters governing the shape of the posterior approximation. This approach is known as \textit{structured} or \textit{fixed-form} Variational Bayes~\citep{honkela,Storkey00dynamictrees,saul1996exploiting}. Usually, the posterior approximation is chosen to be a specific member of the exponential family of distributions:
\begin{equation}
\label{eq:expfam}
q_{\eta}(x) = \exp[T(x)\eta - U(\eta)]\nu(x),
\end{equation}
where $T(x)$ is a $1 \times k$ vector of sufficient statistics, $U(\eta)$ takes care of normalization, and $\nu(x)$ is a base measure. The $k \times 1$ vector $\eta$ is often called the set of \textit{natural parameters} of the exponential family distribution $q_{\eta}(x)$. Using this approach, the variational optimization problem in (\ref{eq:post2}) reduces to a parametric optimization problem in $\eta$:
\begin{equation}
\label{eq:oursol}
\hat{\eta} = \arg\min_{\eta} \E_{q_{\eta}(x)}[ \log q_{\eta}(x) - \log p(x,y)]. 
\end{equation}
If our posterior approximation is of an analytically tractable form, the negative entropy term $\E_{q(x)}[\log q(x)]$ in (\ref{eq:oursol}) can often be evaluated analytically. If we can then also determine $\E_{q(x)}[\log p(x,y)]$ and its derivatives with respect to $\eta$, the optimization problem can be solved using gradient-based optimization or fixed-point algorithms. Posterior approximations of this type are often much more accurate than a factorized approximation, but the requirement of being able to evaluate $\E_{q(x)}[\log q(x)]$ and $\E_{q(x)}[\log p(x,y)]$ analytically is very restrictive. In addition, approximations of this type generally do not allow us to use the fast EM type optimization algorithms often used with factorized approximations \citep[see][Ch. 10]{bishop2006pattern}. In the next section, we draw a parallel between the optimization problem of variational Bayes and linear regression, which allows us to develop a new optimization algorithm that pushes back these limitations significantly.

\section{Variational Bayes as linear regression}
\label{linreg}

For notational convenience we will write our posterior approximation in the adjusted form,
\begin{equation}
\label{eq:unnormalizedq}
\tilde{q}_{\tilde{\eta}}(x) = \exp[\tilde{T}(x)\tilde{\eta}],
\end{equation}
where we have assumed a constant base measure $\nu(x)=1$, and we have replaced the normalizer $U(\eta)$ by adding a constant to the vector of sufficient statistics: $\tilde{T}(x) = (1, T(x))$ and $\tilde{\eta} = (\eta_{0}, \eta')'$. If $\eta_{0}$ is equal to $-U(\eta)$, \eqref{eq:unnormalizedq} describes the same family of (normalized) distribution functions as \eqref{eq:expfam}. If $\eta_{0}$ is different from $-U(\eta)$ then \eqref{eq:unnormalizedq} describes a rescaled (unnormalized) version of this distribution function.
\npar
To work with $\tilde{q}_{\tilde{\eta}}(x)$, we use the unnormalized version of the KL-divergence, which is given by
\begin{align}
\label{eq:klunnorm}
D[\tilde{q}_{\tilde{\eta}}(x) | p(x,y)] & = \int \tilde{q}_{\tilde{\eta}}(x) \log \frac{\tilde{q}_{\tilde{\eta}}(x)}{p(x,y)} dx - \int \tilde{q}_{\tilde{\eta}}(x) dx\\
& = \int \exp[\tilde{T}(x)\tilde{\eta}] [\tilde{T}(x)\tilde{\eta} - \log p(x,y)] dx - \int \exp[\tilde{T}(x)\tilde{\eta}] dx. \nonumber
\end{align}
At the minimum this gives $\eta_{0} = \E_{q}[\log p(x,y) - \log q(x)] - U(\eta)$ as shown in Appendix~\ref{app:normalised}. The other parameters $\eta$ have the same minimum as in the normalized case.
\npar
Taking the gradient of (\ref{eq:klunnorm}) with respect to the natural parameters $\tilde{\eta}$ we have
\begin{align} \label{eq:klgrad}
\nabla_{\tilde{\eta}} D[\tilde{q}_{\tilde{\eta}}(x) | p(x,y)] = 
\int \tilde{q}_{\tilde{\eta}}(x) [\tilde{T}(x)'\tilde{T}(x)\tilde{\eta} - \tilde{T}(x)'\log p(x,y)] dx.
\end{align}
Setting this expression to zero in order to find the minimum gives
\begin{align} \label{eq:vblinreg_int}
\tilde{\eta} = \left[ \int \tilde{q}_{\tilde{\eta}}(x) \tilde{T}(x)'\tilde{T}(x) dx \right]^{-1}\left[\int \tilde{q}_{\tilde{\eta}}(x) \tilde{T}(x)'\log p(x,y) dx\right],
\end{align}
or equivalently
\begin{equation}
\label{eq:vblinreg}
\tilde{\eta} = \E_{q}[\tilde{T}(x)'\tilde{T}(x)]^{-1}\E_{q}[\tilde{T}(x)'\log p(x,y)].
\end{equation}
We have implicitly assumed that the Fisher information matrix, $\E_{q}[\tilde{T}(x)'\tilde{T}(x)]$ is non-singular, which will be the case for any identifiable approximating exponential family distribution $q$. Our key insight is to notice the similarity between \eqref{eq:vblinreg} and the maximum likelihood estimator for linear regression. Recall that in classical linear regression we have that the dependent variable $\{ y_n \in \mathbb{R} : n=1,..,N\}$ is distributed as $N(Y|X\beta,\sigma^2 I)$ where $X$ is the $N \times D$ design matrix, $\beta$ is the $D \times 1$ vector of regression coefficients and $\sigma^2$ is the noise variance. The maximum likelihood estimator for $\beta$ is then
\begin{align} \label{eq:linreg}
 \hat{\beta} = (X'X)^{-1} X'Y.
\end{align}
To see the relation between \eqref{eq:vblinreg} and \eqref{eq:linreg}, associate the design matrix $X$ with the sufficient statistics $\tilde{T}$, the dependent variable $Y$ with the unnormalized log posterior $\log p(x,y)$, and the regression coefficients $\beta$ with the vector of natural parameters $\tilde{\eta}$. If we then consider Monte Carlo estimates of the expectations in \eqref{eq:vblinreg} the analogy is very fitting indeed. A similar analogy is used by \citet{RichardEIS} in the context of importance sampling. Appendix~\ref{eisvb} discusses the connection between their work and ours.
\npar
For notational simplicity, we will assume a constant base measure $\nu(x)=1$ in the remaining discussion, but the linear regression analogy continues to hold if the base measure $\nu(x)$ is non-constant in $x$. In that case, the fixed point condition \eqref{eq:vblinreg} simply becomes
\[
\tilde{\eta} = \E_{q}[\tilde{T}(x)'\tilde{T}(x)]^{-1}\E_{q}[\tilde{T}(x)'(\log p(x,y) - \log \nu(x))],
\]
i.e.\ we perform the linear regression on the residual of the base model $\log \nu(x)$.
\npar
In~\eqref{eq:vblinreg}, unlike~\eqref{eq:linreg}, the right-hand side depends on the unknown parameters, $\eta$. This means that~\eqref{eq:vblinreg} in itself does not constitute a solution to our variational optimization problem. In the next section, we introduce a stochastic approximation algorithm to perform this optimization, without requiring the expectations $\E_{q}[\tilde{T}(x)'\tilde{T}(x)]$ and $\E_{q}[\tilde{T}(x)'\log p(x,y)]$ to be computable analytically. This allows us to extend the fixed-form Variational Bayes approach to situations in which it was previously not applicable. The only requirements we impose on $\log p(x,y)$ is that it is given in closed form. The main requirement on $q_{\eta}(x)$ is that we can sample from it. For simplicity, Sections \ref{sa}, \ref{marglike} and \ref{simul} will also assume that $q_{\eta}(x)$ is in the exponential family. Section \ref{mix} will then show how we can extend this to include mixtures of exponential family distributions. By using these mixtures and choosing $q_{\eta}(x)$ to be of a rich enough type, we can in principle make our approximation arbitrarily precise.

\section{A stochastic approximation algorithm}
\label{sa}

The link between variational Bayes and linear regression in itself is interesting, but it does not yet provide us with a solution to the variational optimization problem of~\eqref{eq:oursol}. We propose solving this optimization problem by viewing~\eqref{eq:vblinreg} as a fixed point update. Let $C = \E_{q}[\tilde{T}(x)'\tilde{T}(x)]$ and $g=\E_{q}[\tilde{T}(x)'\log p(x,y)]$ so that (\ref{eq:vblinreg}) can be written $\tilde{\eta}=C^{-1}g$. We iteratively approximate $C$ and $g$ by weighted Monte Carlo, drawing a single sample $x_{t}^{*}$ from the current posterior approximation $q_{\eta_t}(x)$ at each iteration $t$, and using the update equations
\begin{align}
g_{t+1} &= (1-w)g_{t} + w\hat{g}_{t} \nonumber \\
C_{t+1} &= (1-w)C_{t} + w\hat{C}_{t} \label{eq:cgupdate}
\end{align}
for some $w \in [0,1]$ where $\hat{g}_{t} = \tilde{T}(x_{t}^{*})'\log p(x_{t}^{*},y)$ and $\hat{C}_{t} = \tilde{T}(x_{t}^{*})'\tilde{T}(x_{t}^{*})$. Equation~\ref{eq:cgupdate} downweights earlier iterations when $q$ was less accurate. The parameters are updated as $\tilde{\eta}_{t+1}=C_{t+1}^{-1}g_{t+1}$. $w$ is chosen to be small enough to ensure convergence of the algorithm. Pseudocode is shown in Algorithm~\ref{algo:linregvb}. 

\begin{algorithm}[H]
\caption{Stochastic Optimization for Fixed-Form Variational Bayes}
\label{algo:linregvb}
\begin{algorithmic}

\REQUIRE An unnormalized posterior distribution $p(x,y)$
\REQUIRE A type of approximating posterior $q_{\eta}(x)$
\REQUIRE The total number of iterations $N$

\STATE Initialize $\tilde{\eta}_{1}$ to a first guess, for example by matching the prior $p(x)$
\STATE Initialize $C_{1} = \E_{q_{\eta_{1}}}[\tilde{T}(x)'\tilde{T}(x)]$, or a diagonal approximation of this matrix
\STATE Initialize $g_{1} = C_{1}\tilde{\eta}_{1}$
\STATE Initialize $\bar{C} = \mathbf{0}$
\STATE Initialize $\bar{g} = \mathbf{0}$
\STATE Set step-size $w = 1/\sqrt{N}$

\FOR{$t = 1:N$}

\STATE Simulate a draw $x_{t}^{*}$ from the current approximation $q_{\eta_{t}}(x)$
\STATE Set $\hat{g}_{t} = \tilde{T}(x_{t}^{*})'\log p(x_{t}^{*},y)$, or another unbiased estimate of $\E_{q_{\eta_{t}}}[\tilde{T}(x)'\log p(x,y)]$
\STATE Set $\hat{C}_{t} = \tilde{T}(x_{t}^{*})'\tilde{T}(x_{t}^{*})$, or another unbiased estimate of $\E_{q_{\eta_{t}}}[\tilde{T}(x)'\tilde{T}(x)]$

\STATE Set $g_{t+1} = (1-w)g_{t} + w\hat{g}_{t}$
\STATE Set $C_{t+1} = (1-w)C_{t} + w\hat{C}_{t}$

\STATE Set $\tilde{\eta}_{t+1} = C_{t+1}^{-1}g_{t+1}$

\IF{$t > N/2$}

\STATE Set $\bar{g} = \bar{g} + \hat{g}_{t}$
\STATE Set $\bar{C} = \bar{C} + \hat{C}_{t}$

\ENDIF

\ENDFOR

\RETURN $\hat{\eta} = \bar{C}^{-1}\bar{g}$

\end{algorithmic}
\end{algorithm}

Algorithm~\ref{algo:linregvb} is inspired by a long line of research on stochastic approximation, starting with the seminal work of \citet{robbinsmonro}. Up to first order it can be considered a relatively standard stochastic gradient descent algorithm. At each iteration we have $\tilde{\eta}_{t} = C_{t}^{-1}g_{t}$, which we then update to
\[
\tilde{\eta}_{t+1} = C_{t+1}^{-1}g_{t+1} = [(1-w)C_{t} + w\hat{C}_{t}]^{-1}[(1-w)g_{t} + w\hat{g}_{t}] = [C_{t} + \lambda\hat{C}_{t}]^{-1}[g_{t} + \lambda\hat{g}_{t}],
\]
where $\hat{g}_{t}$ and $\hat{C}_{t}$ are the stochastic estimates generated during iteration $t$, $w$ is the step-size in our algorithm, and $\lambda = w/(1-w)$ is the effective step-size as it is usually defined in the stochastic approximation literature. To characterize this update for small values of $\lambda$ we perform a first order Taylor expansion of $\tilde{\eta}_{t+1}$ around $\lambda=0$, which gives
\begin{equation}
\label{eq:regAsDescent}
\tilde{\eta}_{t+1} = \tilde{\eta}_{t} - \lambda C_{t}^{-1}(\hat{C}_{t}\tilde{\eta}_{t} - \hat{g}_{t}) + \mathcal{O}(\lambda^{2}).
\end{equation}
Comparison with \eqref{eq:klgrad} shows that the stochastic term in this expression ($\hat{C}_{t}\tilde{\eta}_{t} - \hat{g}_{t}$) is an unbiased estimate of the gradient of the KL-divergence $D[q_{\eta_{t}}(x) | p(x,y)]$. Up to first order, the update equation in \eqref{eq:regAsDescent} thus represents a stochastic gradient descent step, pre-conditioned with the $C_{t}^{-1}$ matrix. Since this pre-conditioner is independent of the stochastic gradient approximation at iteration $t$, this gives a valid adaptive stochastic gradient descent algorithm, to which all the usual convergence results apply \citep[see e.g.][]{amaring}.
\npar
If we take small steps, the pre-conditioner $C_{t}^{-1}$ in (\ref{eq:regAsDescent}) will be close to the Riemannian metric $\E_{q_t} \hat{C}_{t} = \E_{q_t}[\tilde{T}(x)'\tilde{T}(x)]$ used in natural gradient descent algorithms like that of \citet{honkela} and \citet{hoffman2012stochastic}. For certain exponential family distributions this metric can be calculated analytically, which would suggest performing stochastic natural gradient descent optimization with updates of the form
\[
\tilde{\eta}_{t+1} = \tilde{\eta}_t - \lambda \left(\tilde{\eta}_t - \E_{q_t}[\tilde{T}(x)'\tilde{T}(x)]^{-1} [\tilde{T}(x^{*})'\log p(x^{*},y)]\right),
\]
where the $\E_{q_t}[\tilde{T}(x)'\log p(x,y)]$ term is approximated using Monte Carlo, but the pre-conditioner $\E_{q_t}[\tilde{T}(x)'\tilde{T}(x)]$ is calculated analytically. At first glance, our approach of approximating $\E_{q_t}[\tilde{T}(x)'\tilde{T}(x)]$ using Monte Carlo only seems to add to the randomness of the gradient estimate, and using the same random numbers to approximate both $\E_{q_t}[\tilde{T}(x)'\log p(x,y)]$ and $\E_{q_t}[\tilde{T}(x)'\tilde{T}(x)]$ leads to biased pre-conditioned gradient approximations at that (although that bias disappears as $\lambda \rightarrow 0$). However, it turns out that approximating both terms using the same random draws increases the efficiency of our algorithm dramatically. The reason for this is analogous to the reason for why the optimal estimator in linear regression is given by $(X'X)^{-1}X'y$ and not $\E[X'X]^{-1}X'y$: by using the same randomness for both the $X'X$ and $X'y$ terms, a large part of the noise in their product cancels out.
\npar
A particularly interesting example of this is when the true posterior distribution is of the same functional form as its approximation, say $p(x,y) = \exp[\tilde{T}(x)\xi]$, in which case Algorithm~\ref{algo:linregvb} will recover the true posterior exactly in $2(k+1)$ iterations, with $k$ the number of sufficient statistics in $q$ and $p$. Assuming the last $k+1$ samples $x_{t}^{*}, t=k+2,...,2k+2$ generated by our algorithm are unique (which holds almost surely for continuous distributions $q$), we have
\begin{eqnarray}
\label{eq:exactconvergence}
\hat{\eta} & = & \left(\sum_{t=k+2}^{2k+2} \tilde{T}(x_{t}^{*})'\tilde{T}(x_{t}^{*})\right)^{-1} \sum_{t=k+2}^{2k+2} \tilde{T}(x_{t}^{*})'\log[p(x_{t}^{*},y)] \nonumber\\
& = & \left(\sum_{t=k+2}^{2k+2} \tilde{T}(x_{t}^{*})'\tilde{T}(x_{t}^{*})\right)^{-1} \sum_{t=k+2}^{2k+2} \tilde{T}(x_{t}^{*})'\tilde{T}(x_{t}^{*})\xi = \xi.
\end{eqnarray}
If the algorithm is run for additional iterations after the true posterior is recovered, the approximation will not change. This is to be contrasted with other stochastic gradient descent algorithms which have non-vanishing variance for a finite number of samples, and is due to the fact that our regression in itself is \textit{noise free}: only its support points are stochastic. This exact convergence will not hold for cases of actual interest, where $p$ and $q$ will not be of the exact same functional form, but we generally still observe a dramatic improvement when using Algorithm~\ref{algo:linregvb} instead of more conventional stochastic gradient descent algorithms. A deeper analysis of the variance of our stochastic approximation is given in Appendix~\ref{sec:choosing}.
\npar
Contrary to most applications in the literature, Algorithm~\ref{algo:linregvb} uses a fixed step size $w = 1/\sqrt{N}$ rather than a declining one in updating our statistics. The analyses of \citet{robbinsmonro} and \citet{amaring} show that a sequence of learning rates $w_{t} = c t^{-1}$ is asymptotically efficient in stochastic gradient descent as the number of iterations $N$ goes to infinity, but this conclusion rests on strong assumptions on the functional form of the objective function (e.g.\ strong convexity) that are not satisfied for the problems we are interested in. Moreover, with a finite number of iterations, the effectiveness of a sequence of learning rates that decays this fast is highly dependent on the proportionality constant $c$. If we choose $c$ either too low or too high, it may take a very long time to reach the efficient asymptotic regime of this learning rate sequence.
\npar
\citet{robustsa} show that a more robust approach is to use a constant learning rate $w = 1/\sqrt{N}$ and that this is optimal for finite $N$ without putting stringent requirements on the objective function. In order to reduce the variance of the last iterate with this non-vanishing learning rate, they propose to use an average of the last $L$ iterates as the final output of the optimization algorithm. The value of $L$ should grow with the total number of iterations, and is usually chosen to be equal to $N/2$. Remarkably, they show that such an averaging procedure can match the asymptotic efficiency of the optimal learning sequence $w_{t} = c t^{-1}$.
\npar
For our particular optimization problem we have observed excellent results using constant learning rate $w = 1/\sqrt{N}$, and averaging starting half-way into the optimization. We perform this averaging on the statistics $g$ and $C$, rather than on the parameters $\tilde{\eta} = C^{-1}g$, which is necessary to remove the bias caused by forming $g$ and $C$ using the same random numbers. As previously described, using this set-up $g_{t}$ and $C_{t}$ are actually weighted MC estimates where the weight of the $j$-th MC sample during the $t$-th iteration ($j \leq t$) is given by $w (1-w)^{t-j}$. Since $w \in (0,1)$, this means that the weight of earlier MC samples declines as the algorithm advances, which is desirable since we expect $q$ to be closer to optimal later in the algorithm's progression.
\npar
If the initial guess for $\tilde{\eta}$ is very far from the optimal value, or if the number of steps $N$ is very small, it can sometimes occur that the algorithm proposes a new value for $\tilde{\eta}$ that does not define a proper distribution, for example because the proposed $\tilde{\eta}$ value corresponds to a negative variance. This is a sign that the number of iterations should be increased: since our algorithm becomes a pre-conditioned gradient descent algorithm as the number of steps goes to infinity, the algorithm is guaranteed to converge if the step size is small enough. In addition, the exact convergence result presented in \eqref{eq:exactconvergence} suggests that divergence is very unlikely if $q_{\eta}(x)$ and $p(x,y)$ are close in functional form: choosing a good approximation will thus also help to ensure fast convergence. Picking a good first guess for $\tilde{\eta}$ also helps the algorithm to converge more quickly. For very difficult cases it might therefore be worthwhile to base this guess on a first rough approximation of the posterior, for example by choosing $\tilde{\eta}$ to match the curvature of $\log p(x,y)$ at its mode. For all our applications we found that a simple first guess for $\tilde{\eta}$ and a large enough number of iterations was sufficient to guarantee a stable algorithm. Our default implementation of Algorithm~\ref{algo:linregvb} is therefore to initialize $\tilde{\eta}$ to (an approximation of) the prior, and to increase the number of iterations until the algorithm is sufficiently stable.
\npar
Like other optimization algorithms for Variational Bayes, Algorithm~\ref{algo:linregvb} will only find a \textit{local} minimum of the KL-divergence. This is generally not a problem when approximating unimodal posterior distributions, such as with the examples in this paper, since the optimization problem then often only has a single optimum \citep[depending on the type of approximation, see][Ch. 10]{bishop2006pattern}. If the true posterior distribution is multimodal and the approximation is unimodal, however, the variational approximation will tend to pick one of the posterior modes and ignore the others \citep{epdivergence}. Although this is often desirable \citep[see e.g.][]{stern09}, there is no guarantee that the recovered local minimum of the KL-divergence is then also a global minimum. %In these cases, one possible strategy is to run Algorithm~\ref{algo:linregvb} multiple times using different starting points, and then to pick the best solution.

\subsubsection{Example: Fitting an exponential distribution}
\begin{SCfigure}[1.0][hb]%[htb]
	\centering
		\includegraphics[width=0.6\textwidth]{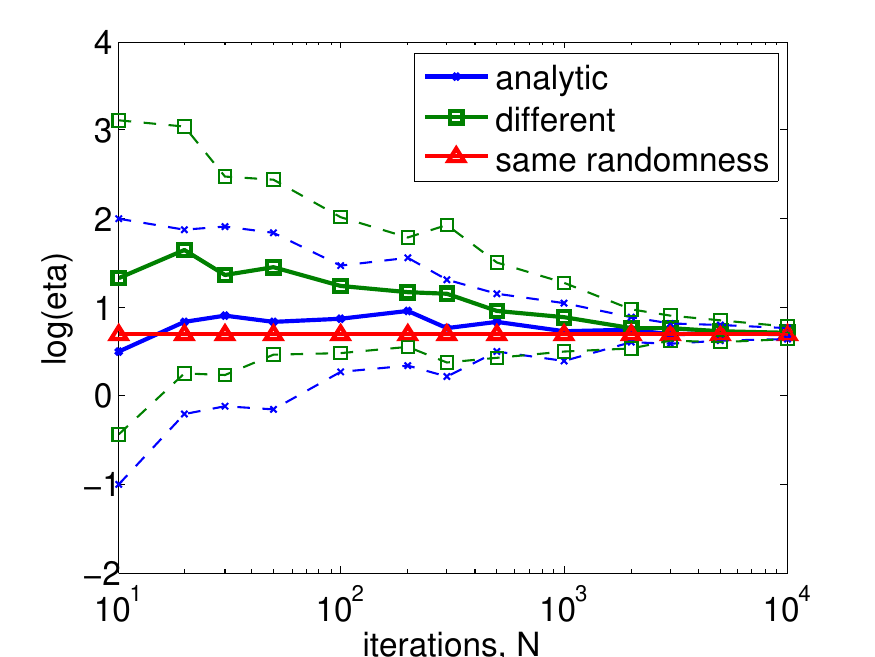}
	\caption{Comparing alternative methods for estimating $\hat{g}_t$ and $\hat{C}_t$ in Algorithm~\ref{algo:linregvb} on a toy example: approximating an exponential posterior with an approximation of the same functional form. Solid lines show the means of the recovered parameters over $100$ repeat runs, and dashed lines show $\pm$ one standard deviation. Using the same random draw to estimate $\hat{g}_t$ and $\hat{C}_t$ (our proposed method) gives exact convergence in $N=4$ iterations}
	\label{fig:lr_exp}
\end{SCfigure}

It is instructive to consider a toy example: approximating an exponential distribution $p(x) = \lambda e^{-\lambda x}$ with a variational approximation of the same functional form. We assume that we are unaware that $p$ happens to be normalized. Our variational approximation has $\tilde{T}= [1, x]$ and rate $\eta$, i.e.\ $q(x)=\eta e^{-\eta x}$. Since the functional form of the variational posterior matches the true posterior, \eqref{eq:exactconvergence} holds and Algorithm~\ref{algo:linregvb} will recover $\eta$ to machine precision in just $2(k+1)=4$ iterations. We contrast this with the performance if two different strategies are used to estimate $\hat{g}_t$ and $\hat{C}_t$ in Algorithm~\ref{algo:linregvb}: i) a different random draw $x^*$ is used for $\hat{g}_t$ and $\hat{C}_t$, ii) $\hat{C}_t$ is calculated analytically using
\begin{align} \label{eq:analyticC}
 \E_{q}[\tilde{T}(x)'\tilde{T}(x)] = \left[ \begin{array}{cc} 
                                            1 & -\eta^{-1}  \\
					    -\eta^{-1} & \eta^{-2} 
                                           \end{array} \right]. 
\end{align}
These seemingly similar alternatives perform dramatically worse than Algorithm~\ref{algo:linregvb}. We set the true $\lambda := 2$, and initialize $\eta := 1$ and $C := I_2$, the identity matrix. Figure~\ref{fig:lr_exp} shows the mean and variance of the estimates of $\log(\eta)$ across $100$ repeat runs of each method with varying number of iterations $N$. We see it takes option i (``different randomness'') and ii (``analytic'') well over $1000$ iterations to give a reasonable answer, and even with $N=10^4$ samples, option i) estimates $\hat{\eta} = 2.04 \pm 0.15 $ and option ii) $ 2.01 \pm 0.11 $.

\section{Marginal likelihood and approximation quality}
\label{marglike}

The stochastic approximation algorithm presented in the last section serves to minimize the Kullback-Leibler divergence between $q_{\eta}(x)$ and $p(x|y)$, given by
\[
D(q_{\eta}|p) = \E_{q_{\eta}}\left[\log \frac{q_{\eta}(x)}{p(x|y)}\right] = \E_{q_{\eta}}\left[ \log \frac{q_{\eta}(x)}{p(x,y)}\right] + \log p(y).
\]
As discussed before, we do not need to know $p(y)$ (the marginal likelihood)  in order to \emph{minimize} $D(q_{\eta}|p)$ as $p(y)$ does not depend on $\eta$, but we do need to know it if we want to determine the quality of the approximation, as measured by the final Kullback-Leibler divergence. In addition, the constant $p(y)$ is also essential for performing Bayesian model comparison or model averaging. This section presents a method for approximating the marginal likelihood and final Kullback-Leibler divergence.
\npar
When our algorithm has converged, we have the following identity
\begin{equation}
\label{eq:approxconv}
\log p(x,y) = \hat{\eta}_{0} + T(x)\eta + r(x) = \hat{\eta}_{0} + U(\eta) + \log q_{\eta}(x) + r(x),
\end{equation}
where $r(x)$ is the `residual' or `error term' in the linear regression of $\log p(x,y)$ on the sufficient statistics of $q_{\eta}(x)$, and where $U(\eta)$ is the normalizer of $q_{\eta}(x)$. The intercept of the regression is
\[
\hat{\eta}_{0} = \E_{q_{\eta}}\left[ \log p(x,y) - \log q_{\eta}(x) \right] - U(\eta),
\] 
where $\E_{q_{\eta}}\left[ \log p(x,y) - \log q_{\eta}(x) \right] = \hat{\eta}_{0} + U(\eta)$ may be recognized as the usual VB lower bound on the log marginal likelihood. Exponentiating (\ref{eq:approxconv}) yields
\[
p(x,y) = \exp[\hat{\eta}_{0} + U(\eta)] q_{\eta}(x) \exp(r(x)),
\]
which we need to integrate with respect to $x$ in order to find the marginal likelihood $p(y)$. Doing so gives
\begin{equation}
\label{eq:marglike}
p(y) = \exp[\hat{\eta}_{0} + U(\eta)] \E_{q_{\eta}}[\exp(r(x))].
\end{equation}
At convergence we have that $\E_{q_{\eta}}[r(x)] = 0$. Jensen's inequality then tells us that
\[
\E_{q_{\eta}}[\exp(r(x))] \geq 1,
\]
so that $\hat{\eta}_{0} + U(\eta)$ is indeed a lower bound on the log marginal likelihood. If our approximation is perfect, the KL-divergence is zero and $r(x)$ is zero almost everywhere. In that case the residual term vanishes and the lower bound will be tight, otherwise it will underestimate the true marginal likelihood. The lower bound $\hat{\eta}_{0} + U(\eta)$ is often used in model comparison, which works well if the KL-divergence between the approximate and true posterior distribution is of approximately the same size for all models that are being compared. However, if we compare two very different models this will often not be the case, and the model comparison will be biased as a result. In addition, as opposed to the exact marginal likelihood, the lower bound gives us no information on the quality of our posterior approximation. It would therefore be useful to obtain a better estimate of the marginal likelihood.
\npar
One approach to doing this would be to evaluate the expectation in~\eqref{eq:marglike} using Monte Carlo sampling. Some analysis shows that this corresponds to approximating $p(y)$ using importance sampling, with $q_{\eta}(x)$ as the candidate distribution. It is well known that this estimator of the marginal likelihood may have infinite variance, unless $r(x)$ is bounded from above \citep[see e.g.][p. 114]{gewekebook}. In general, we cannot guarantee the boundedness of $r(x)$ for our approach, so we will instead approximate the expectation in~\eqref{eq:marglike} using something that is easier to calculate.
\npar
At convergence, we know that the mean of $r(x)$ is zero when sampling from $q_{\eta}(x)$. The variance of $r(x)$ can be estimated using the mean squared error of the regressions we perform during the optimization, with relatively low variance. We denote our estimate of this variance by $s^{2}$. The assumption we will then make in order to approximate $\log p(y)$ is that $r(x)$ is approximately distributed as a normal random variable with these two moments. This leads to the following simple estimate of the log marginal likelihood
\begin{equation}
\label{eq:newmarglikeapprox}
\log p(y) \approx \hat{\eta}_{0} + U(\eta) + \frac{1}{2}s^{2}.
\end{equation}
That is, our estimate of the marginal likelihood is equal to its lower bound plus a correction term that captures the error in our posterior approximation $q_{\eta}(x)$. Similarly, we can approximate the KL-divergence of our posterior approximation as
\[
D(q_{\eta}|p) \approx \frac{1}{2}s^{2}.
\]
The KL-divergence is approximately equal to half the mean squared error in the regression of $\log p(x,y)$ on the sufficient statistics of the approximation. This relationship should not come as a surprise: this mean squared error is exactly what we minimize when we perform linear regression.
\npar
The scale of the KL-divergence is highly dependent on the amount of curvature in $\log p(x|y)$ and is therefore not easily comparable across different problems. If we scale the approximate KL-divergence to account for this curvature, this naturally leads to the R-squared measure of fit for regression modeling:
\[
R^{2} = 1 - \frac{s^{2}}{\var_{q}[\log p(x,y)]}.
\]
The R-squared measure corrects for the amount of curvature in the posterior distribution and is therefore comparable across different models and data sets. In addition it is a well-known measure and easily interpretable. We therefore propose to use the R-squared as the measure of approximation quality for our variational posterior approximations. Although we find the R-squared to be a useful measure for the majority of applications, it is important to realize that it mostly contains information about the mass of the posterior distribution and its approximation, and not directly about their moments. It is therefore possible to construct pathological examples in which the R-squared is relatively high, yet the (higher) moments of the posterior and its approximation are quite different. This may for example occur if the posterior distribution has very fat tails.
\npar
Section \ref{mixture_example} provides an application of the methods developed here. In that section, Figure~\ref{fig:mixkldiv} shows that the approximation of the KL-divergence is quite accurate, especially when the approximation $q_{\eta}(x)$ is reasonably good. The same figure also shows that the approximation of the marginal likelihood proposed here \eqref{eq:newmarglikeapprox} is much more accurate than the usual lower bound. In Sections \ref{simul} and \ref{mix}, we also calculate the R-squared measure of approximation quality for a number of different posterior approximations, and we conclude that it corresponds well to visual assessments of the approximation accuracy.
\npar
The discussion up to this point represents the core ideas of this paper. To make our approach more general and computationally efficient we now provide a number of extensions in two separate sections. Section~\ref{simul} discusses modifications of our stochastic approximation algorithm to improve efficiency, and Section~\ref{mix} generalizes the exponential family approximations $q(x)$ used so far to include mixtures of exponential family distributions. Examples are given throughout. Finally, Section~\ref{conclusion} concludes.

\section{Extensions I: Improving algorithmic efficiency}
\label{simul}

Algorithm~\ref{algo:linregvb} approximates the expectations $\E_{q_{\eta}}[\tilde{T}(x)'\log p(x,y)]$ and $\E_{q_{\eta}}[\tilde{T}(x)'\tilde{T}(x)]$ by simply drawing a sample $x_{t}^{*}$ from $q_{\eta_{t}}(x)$ and using this sample to calculate
\begin{eqnarray}
\hat{g}_{t} & = & \tilde{T}(x_{t}^{*})'\log p(x_{t}^{*},y) \nonumber\\
\hat{C}_{t} & = & \tilde{T}(x_{t}^{*})'\tilde{T}(x_{t}^{*}). \nonumber
\end{eqnarray}
This works remarkably well because, as Section~\ref{sa} explains, using the same random draw $x_{t}^{*}$ to form both estimates, part of the random variation in $\tilde{\eta} = C^{-1}g$ cancels out. However, it is certainly not the only method of obtaining unbiased approximations of the required expectations, and in this section we present alternatives that often work even better. In addition, we also present alternative methods of parameterizing our problem, and we discuss ways of speeding up the regression step of our algorithm.

\subsubsection{Example: Binary probit regression}
\begin{figure}[H]
	\centering
		\includegraphics[width=\textwidth]{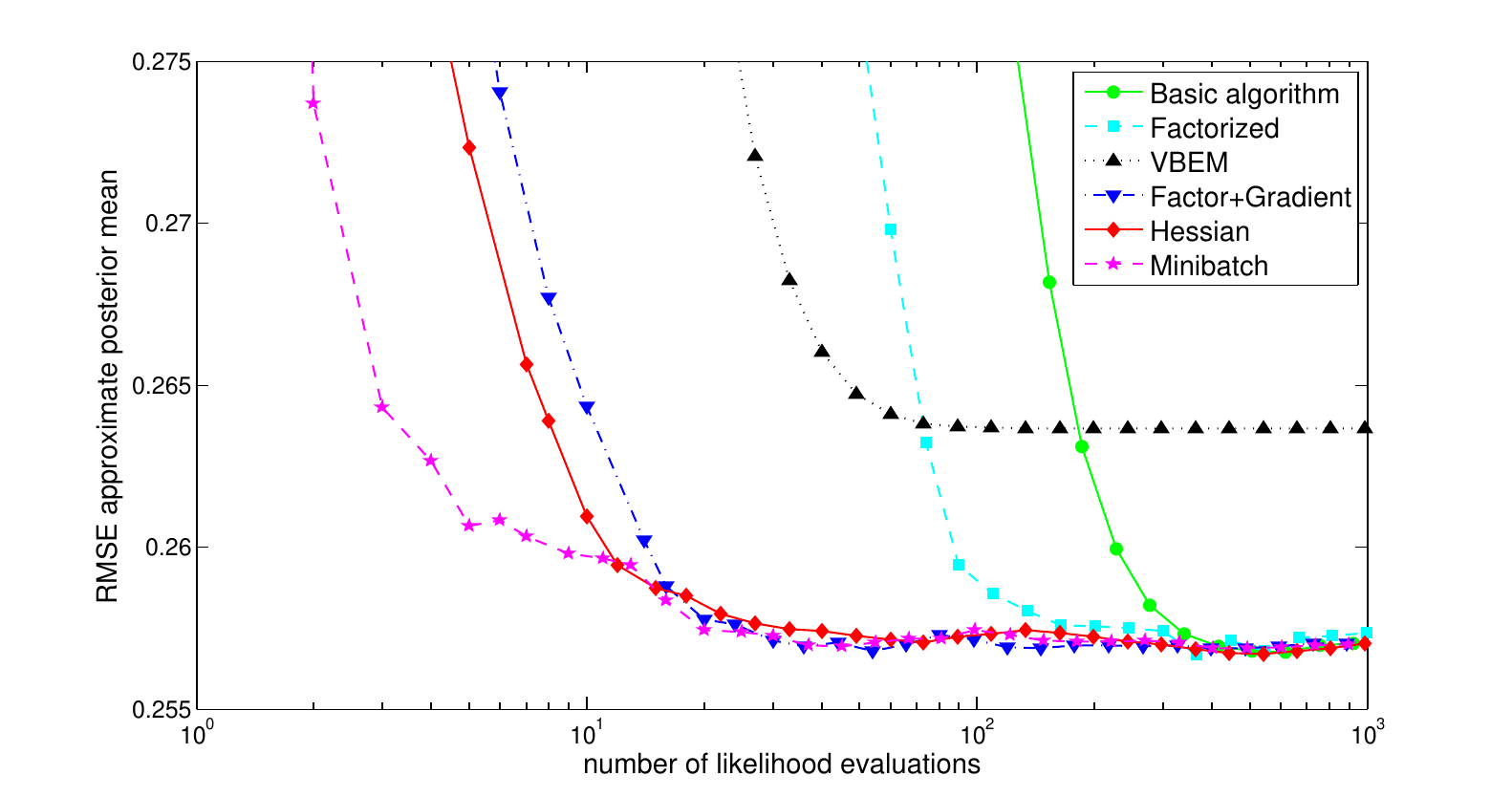}
	\caption{RMSE approximate posterior mean as a function of the number of likelihood evaluations for the different implementations of our algorithm and VBEM. Green: our basic algorithm (Section~\ref{sa}). Cyan: using factor structure (Section~\ref{factorstructure}). Black: the standard VBEM algorithm. Blue: using both factor structure and the gradient of the log posterior (Section~\ref{usinggrad}). Red: using the Hessian of the log posterior with linear transformation for efficiency (Sections~\ref{hess} and \ref{lintrans}). Magenta: using the Hessian, linear transformation and minibatches of data (Section~\ref{sec:subsample}). }
	\label{fig:probitRMSE}
\end{figure}

To illustrate the different versions of our posterior approximation algorithm, we will use binary probit regression as a running example. Binary probit regression is a classic model in statistics, also referred to as binary classification in the machine learning literature. Here we take a Bayesian approach to probit regression to demonstrate the performance of our methodology relative to existing variational approaches. We have $N$ observed data pairs $(y_i \in \{0,1\},v_i \in \mathbb{R}^M)$, and we model $y_i|v_i$ as $P(y_i=1|v_i,x)=\phi(x'v_i)$ where $\phi(.)$ is the standard Gaussian cdf and $x \in \mathbb{R}^M$ is a vector of regression coefficients, for which we assume an elementwise Gaussian prior $N(0,1)$. This is a model for which existing approaches are straightforward so it is interesting to compare their performance to our method. Of course the major benefit of our approach is that it can be applied in a much wider class of models. For all versions of our method the variational approximation used is a full covariance multivariate normal distribution. 
\npar
We use data simulated from the model, with $N=100$ and $M=5$, to be able to show the performance averaged over $500$ datasets and many different settings of the algorithm. We compare our algorithm to the VBEM algorithm of \citet{ormerod:2010} which makes use of the fact that the expectations required for this model can be calculated analytically. We choose not to do this for our method to investigate how effective our MC estimation strategy can be. For completeness we also compare to variational message passing~\citep[VMP,][]{Winn2006}, a message passing implementation of VBEM, and expectation propagation~\citep[EP,][]{MinkaThesis}, which is known to have excellent performance on binary classification problems~\citep{Nickisch2008}. These last two alternatives are both implemented in Infer.NET~\citep{InferNET10} a library for probabilistic inference in graphical models, whereas we implement VBEM and our approximation algorithm ourselves in MATLAB. VMP and VBEM use a different variational approximation to our methods, introducing auxiliary variables $z_i \sim N(x'v_i,1)$, with $z_i$ constrained to be positive if $y_i=1$ and negative otherwise. A factorized variational posterior $q(x)\prod_i q(z_i)$ is used, where $q(x)$ is multivariate normal and each $q(z_i)$ can be thought of as a truncated univariate Gaussian. 
\npar
For all implementations of our algorithm, we initialize the posterior approximation to the prior. All algorithms then use a single random draw to update the parameters during each iteration. This is often not the best implementation in terms of computational efficiency, since the contributions of multiple draws can often be calculated in parallel at little extra cost, and using antithetic sampling (i.e. sampling of negatively correlated draws) can reduce the variance of our approximations. By using the most basic implementation, however, we can more clearly compare the different stochastic approximations proposed in this section. Since the time required to run the different algorithms is strongly dependent on their precise implementation (e.g. the chosen programming language), we choose to perform this comparison by looking at statistical efficiency, as measured by the accuracy as a function of the number of likelihood evaluations, rather than the running time of the algorithms.
\npar
Since this experiment is on synthetic data we are able to assess performance in terms of the method's ability to recover the known regression coefficients $x$, which we quantify as the root mean squared error (RMSE) between the variational mean and the true regression weights, and the ``log score'': the log density of the true weights under the approximate variational posterior. The log score is useful because it rewards a method for finding good estimates of the posterior variance as well as the mean, which should of course be central to any approximate Bayesian method.
\npar
Figure~\ref{fig:probitRMSE} shows the performance of the different versions of our algorithm as presented in the following discussion, as well as the performance of the VBEM algorithm of \citet{ormerod:2010}. As can be seen from this graph, our approximation method achieves a lower RMSE than the VBEM algorithm. This is because of the extra factorization assumptions made by VBEM when introducing the $z_i$ variables. Where the improvement in the RMSE is noticeable, the difference in log score is dramatic: $0.193$ versus $-4.46$ (not shown), indicating that our approximation gives significantly better estimates of the variance than VBEM. The average R-squared obtained by our variational approximation was $0.97$, indicating a close fit to the exact posterior distribution. In terms of accuracy, our results are very similar to those of EP, which obtained an RMSE and log score identical to those of our approximation (up to 3 significant digits). As expected, VMP gave consistent results with VBEM: an RMSE of $0.265$ and a log score of $-4.56$.
As can be seen from Figure~\ref{fig:probitRMSE}, our basic algorithm is considerably slower than VBEM in terms of the number of likelihood evaluations that are required to achieve convergence. In terms of wall clock time, our basic algorithm ran about an order of magnitude slower than VBEM, although it could easily be sped up by using multiple random draws in parallel. The basic algorithm was about as fast as EP and VMP, needing about 15 milliseconds to converge on this small data set, but note that the system set ups were not completely comparable: EP and VMP were run on a laptop rather than a desktop, and Infer.NET is implemented in C\# rather than MATLAB.
\npar
The remainder of this section introduces the other implementations of our variational approximation, presented in Figure~\ref{fig:probitRMSE}, some of which are much faster and more computationally efficient than both our basic algorithm and VBEM.
\FloatBarrier

\subsection{Making use of factor structure}
\label{factorstructure}
For most statistical problems, including our probit regression model, the log posterior can be decomposed into a number of additive factors, i.e.\ $\log p(x,y) = \sum_{j=1}^{N} \log \phi_{j}(x,y)$. The optimality condition in~\eqref{eq:vblinreg} can then also be written as a sum:
\[
\tilde{\eta} = \sum_{j=1}^{N} \E_{q}[\tilde{T}(x)'\tilde{T}(x)]^{-1} \E_{q}[\tilde{T}(x)'\log \phi_{j}(x,y)].
\]
This means that rather than performing one single linear regression we can equivalently perform $N$ separate regressions.
\begin{eqnarray}
\label{eq:sepreg}
\hat{\eta} & = & \sum_{j=1}^{N} \hat{\eta}^{j} \\
\hat{\eta}^{j} & = & \E_{q}[\tilde{T}(x)'\tilde{T}(x)]^{-1} \E_{q}[\tilde{T}(x)'\log \phi_{j}(x,y)].
\end{eqnarray}
One benefit of this is that some of the factors $\phi_{j}(x,y)$ may be conjugate to the posterior approximation, such as the prior $p(x)$ in our probit regression example. The regression coefficients $\hat{\eta}^{j}$ for these conjugate factors are known analytically and do not need to be approximated. 
\npar
More importantly, the separate coefficients $\hat{\eta}^{j}$ in \eqref{eq:sepreg} can often be calculated using regressions of much lower dimension than the full vector of natural parameters since the factors $\phi_{j}(x,y)$ often only depend on a few of the sufficient statistics of our approximation. This occurs when the factors are functions of low dimensional projections or subsets of $x$. For example, we might have $\phi_{j}(x,y)=\phi_{j}(x_{R},y)$, where $x_{R}$ contains a subset of the variables in $x$. In that case, it follows from the properties of the exponential family that $\log \phi_{j}(x,y)$ will have zero partial correlation with all the sufficient statistics in $\tilde{T}(x)$, after controlling for the sufficient statistics of the marginal $q(x_{R})$ \citep[see][Section 5.5]{wainwright2008graphical}. In other words, we have
\[
\log \phi_{j}(x,y) = \tilde{T}_{R}(x)\hat{\eta}^{j}_{R}  + r(x), \text{ with } \E_{q}[\tilde{T}(x)'r(x)] = 0,
\]
where $\tilde{T}_{R}(x)$ is that subset of the statistics in $\tilde{T}(x)$ that is sufficient for $q(x_{R})$, and $\hat{\eta}^{j}_{R}$ is the corresponding subset of the parameters in $\hat{\eta}^{j}$. The `residual' $r(x)$ is orthogonal to the remaining sufficient statistics, i.e. the factor $\log \phi_{j}(x,y)$ has zero partial correlation to the sufficient statistics that are not in the subset $\tilde{T}_{R}(x)$, which means that the coefficients of those statistics will be zero. Statistics that are known to have a zero coefficient can of course be omitted from the regression, leading to the low dimensional regression
\[
\hat{\eta}^{j}_{R} = \E_{q}[\tilde{T}_{R}(x)'\tilde{T}_{R}(x)]^{-1} \E_{q}[\tilde{T}_{R}(x)'\log \phi_{j}(x,y)].
\]
By performing these lower dimensional regressions we can reduce the variance of the stochastic approximation algorithm, as well as reduce the overhead needed to store and invert $C = \E_{q}[\tilde{T}(x)'\tilde{T}(x)]$.
\npar
Our probit regression model provides a straightforward example, for which the log joint density of $x$ and $y$ has the following factor structure
\[
\log p(x,y) = \log p(x) + \sum_{i=1}^N \log p(y_{i}|v_{i},x).
\]
Here, each likelihood factor $p(y_{i}|v_{i},x)$ depends on all the parameters $x$, but only through the univariate product $f_{i} = x'v_{i}$. We can emphasize this by writing our model as
\[
\log p(x,y) = \log p(x) + \sum_{i=1}^N \log p(y_{i}|f_{i}),
\]
where the new variables $f_{i}$ are linked to the parameters $x$ through the relationship $f_{i} = x'v_{i}$. When we sample $x$ from its multivariate normal approximate posterior, the resulting $f_{i}$'s will have univariate normal distributions $q_{\eta}(f_{i}) = N[\mu_{i},\sigma^{2}_{i}]$, with $\mu_{i} = v_{i}'\E_{q}[x]$ and $\sigma^{2}_{i} = v_{i}'\Var_{q}[x]v_{i}$. This means that the factors $\log p(y_{i}|f_{i})$ will have zero partial correlation to the statistics $\tilde{T}(x)$ after controlling for the sufficient statistics of the marginals $q_{\eta}(f_{i})$, being $f_{i}$ and $-0.5f_{i}^{2}$. Approximating $p(x|y)$ by a multivariate Gaussian is thus equivalent to approximating the likelihood factors $p(y_{i}|f_{i})$ by univariate Gaussian likelihood terms in $f_i$. Using this, we can write our unnormalized approximate posterior $\tilde{q}_{\tilde{\eta}}(x)$ as
\begin{eqnarray}
\log \tilde{q}_{\tilde{\eta}}(x) & = & \log p(x) + \sum_{i=1}^N \left[ \tilde{\eta}_{i,0} + \tilde{\eta}_{i,1}f_{i} - 0.5\tilde{\eta}_{i,2}f_{i}^{2} \right]\\
& = & \log p(x) + \sum_{i=1}^N \left[ \tilde{\eta}_{i,0} + \tilde{\eta}_{i,1}x'v_{i} - 0.5\tilde{\eta}_{i,2}(x'v_{i})^{2} \right] \nonumber
\end{eqnarray}
where $\tilde{\eta}_{i,0}$, $\tilde{\eta}_{i,1}$, and $\tilde{\eta}_{i,2}$ are the natural parameters of the univariate Gaussian approximation of the likelihood term $p(y_{i}|f_{i})$.
These parameters can now be optimized by performing a separate regression for each likelihood factor, using the statistics
\[
\tilde{T}(f_{i})' = \left[\begin{array}{c} 1 \\ f_{i} \\ -0.5f_{i}^{2} \end{array}\right] = \left[\begin{array}{c} 1 \\ v_{i}'x \\ -0.5(v_{i}'x)^{2} \end{array}\right],
\]
and regressing these against the likelihood factors $\log p(y_{i}|v_{i},x_{i})$. At each iteration of Algorithm~\ref{algo:linregvb}, we can then update the natural parameters of each approximate likelihood term using
\begin{eqnarray}
\hat{g}_{t,i} & = & \tilde{T}(v_{i}'x_{t}^{*})'\log[p(y_{i}|v_{i},x_{t}^{*})] \\
\hat{C}_{t,i} & = & \tilde{T}(v_{i}'x_{t}^{*})'\tilde{T}(v_{i}'x_{t}^{*}) \nonumber \\
g_{t+1,i} & = & (1-w)g_{t,i} + w\hat{g}_{t,i} \nonumber \\
C_{t+1,i} & = & (1-w)C_{t,i} + w\hat{C}_{t,i} \nonumber \\
\tilde{\eta}_{t+1,i} & = & C_{t+1,i}^{-1}g_{t+1,i}. \nonumber
\end{eqnarray}
Rather than performing a single regression of dimension $1 + M(M+3)/2$, we may thus equivalently perform $N$ regressions of dimension 3. Performing these lower dimensional regressions is computationally more efficient as long as $N$ is not very large, and it is also statistically more efficient. Figure~\ref{fig:probitRMSE} shows that this factorized regression implementation of our approximation indeed needs far fewer random draws to achieve convergence. All $N$ regressions can be performed in parallel, which offers further opportunities for computational gain on multicore machines or computer clusters. 
\npar
So far, we have assumed that we sample $x^{*}$ and then form the $f_{i}$ by multiplying with the $v_{i}$, but note that we can equivalently sample the $f_{i}$ directly and separately from their univariate Gaussian approximate posteriors $q_{\eta}(f_{i}) = N[\mu_{i}(\eta,v_{i}),\sigma^{2}_{i}(\eta,v_{i})]$. For the current example we find that both implementations are about equally efficient.

\subsection{Using the gradient of the log posterior}
\label{usinggrad}
Using the Frisch-Waugh-Lovell theorem \citep{lovell2008simple}, we can remove the constant from the sufficient statistics $\tilde{T}(x)$ and rewrite the optimality condition (\ref{eq:vblinreg}) in its normalized form (this is shown for our particular application in Appendix~\ref{app:normalised}):
\begin{equation}
\label{eq:covgrad}
\hat{\eta} = \Cov_{q}[T(x),T(x)]^{-1}\Cov_{q}[T(x),\log p(x,y)].
\end{equation}
Furthermore, using the properties of the exponential family of distributions, we know that
\begin{equation}
\label{eq:reggrad1}
\Cov_{q}[T(x),T(x)] = \nabla_{\eta} \E_{q_{\eta}}[T(x)],
\end{equation}
which we take to denote the transposed Jacobian matrix of $\E_{q_{\eta}}[T(x)]$, and
\begin{equation}
\label{eq:reggrad2}
\Cov_{q}[T(x),\log p(x,y)] = \nabla_{\eta} \E_{q_{\eta}}[\log p(x,y)],
\end{equation}
which denotes the column vector gradient of $\E_{q_{\eta}}[\log p(x,y)]$ (since $p(x,y)$ is a scalar valued function).
\npar
Both $\E_{q_{\eta}}[T(x)]$ and $\E_{q_{\eta}}[\log p(x,y)]$ can be approximated without bias using Monte Carlo. By differentiating these Monte Carlo approximations we can then obtain unbiased estimates of their derivatives. This is easy to do as long as the pseudo-random draw $x^{*}$ from $q_{\eta}(x)$ is a differentiable function of the parameters $\eta$, given our random number seed $z^{*}$:
\begin{eqnarray}
\label{eq:grad}
x^{*} & = & s(\eta,z^{*}), \text{ with $s()$ and $z^{*}$ such that }  x^{*} \sim q_{\eta}(x) \\
\hat{g} & = & \nabla_{\eta} \log p(s(\eta,z^{*}),y) = \nabla_{\eta} s(\eta,z^{*}) \nabla_{x} \log p(x^*,y) \nonumber \\
\hat{C} & = &  \nabla_{\eta} T(s(\eta,z^{*}))= \nabla_{\eta} s(\eta,z^{*}) \nabla_{x} T(x^*). \nonumber 
\end{eqnarray}
By using the same random number seed $z^{*}$ in both Monte Carlo approximations we once again get the beneficial variance reduction effect described in Section~\ref{sa}.
\npar
Performing a single iteration using (\ref{eq:grad}) provides about the same information as doing $2\times\dim(x)$ iterations with the basic algorithm, making it more computationally efficient if the gradients can be obtained analytically.
\npar
We can also do updates of this form while still making use of the factor structure of the posterior distribution, as proposed above for the probit regression example. Using this example, and assuming we sample the $f_{i}$ separately (see last paragraph of Section~\ref{factorstructure}), this gives the following regression statistics for each of the $N$ low dimensional regressions:
\begin{align}
f_{i}^{*} & = s_{i}(\eta,z_{i}^{*}) = \mu_{i}(\eta,v_{i}) + \sigma_{i}(\eta,v_{i})z_{i}^{*}, \text{ with } z_{i}^{*} \sim N(0,1) \\
\hat{g}_{i} & = \left[\begin{array}{c} \frac{\partial s_{i}(\eta,z_{i}^{*})}{\partial \eta_{i,1}} \\ \frac{\partial s_{i}(\eta,z_{i}^{*})}{\partial \eta_{i,2}} \end{array}\right] \frac{\partial \log p(y|f^*)}{\partial f_{i}} \nonumber\\
& = \left[\begin{array}{c} \sigma_{i}^{2} \frac{\partial \log p(y_{i}|f_{i}^*)}{\partial f_{i}} \\ (-\mu_{i}\sigma_{i}^{2} - 0.5\sigma_{i}^{3}z_{i}^{*})\frac{\partial \log p(y_{i}|f_{i}^*)}{\partial f_{i}} \end{array}\right] \nonumber \\
\hat{C}_{i} & =  \left[\begin{array}{c} \frac{\partial s_{i}(\eta,z_{i}^{*})}{\partial \eta_{i,1}} \\ \frac{\partial s_{i}(\eta,z_{i}^{*})}{\partial \eta_{i,2}} \end{array}\right] \left[\begin{array}{cc} \frac{\partial T_{i,1}(f_{i}^*)}{\partial f_{i}^*} & \frac{\partial T_{i,2}(f_{i}^*)}{\partial f_{i}^*} \end{array}\right] \nonumber\\
& =  \left[\begin{array}{cc} \sigma_{i}^{2} & -\sigma_{i}^{2}f_{i}^* \\ -\mu_{i}\sigma_{i}^{2} - 0.5\sigma_{i}^{3}z_{i}^{*} & (\mu_{i}\sigma_{i}^{2} + 0.5\sigma_{i}^{3}z_{i}^{*})f_{i}^* \end{array}\right]. \nonumber 
\end{align}
Figure~\ref{fig:probitRMSE} shows the performance of this approximation on our probit example, showing again a large gain in efficiency with respect to the approximations introduced earlier. Empirically, we find that using gradients also leads to more efficient stochastic optimization algorithms for many other applications. For some problems the posterior distribution will not be differentiable in some of the elements of $x$, for example when $x$ is discrete. In that case the stochastic approximations presented here may be combined with the basic approximation of Section~\ref{sa}.
\npar
In addition, for many samplers $\nabla_{\eta} s(\eta,z^{*})$ may be not defined, e.g.\ rejection samplers. However, for the gradient approximations it does not matter what type of sampler is actually used to draw $x^{*}$, only that it is from the correct distribution. A correct strategy is therefore to draw $x^{*}$ using any desired sampling algorithm, and then proceeding as if we had used a different sampling algorithm for which $\nabla_{\eta} s(\eta,z^{*})$ is defined. For example, we might use a nondifferentiable rejection sampler to draw a univariate $x^{*}$, and then calculate (\ref{eq:grad}) as if we had used an inverse-transform sampler, for which we have
\begin{equation}
\label{eq:invtrans}
\frac{\partial}{\partial \eta_{i}} s(\eta,z^{*}) = -\frac{\frac{\partial}{\partial \eta_{i}} Q_{\eta}(x^{*})}{q_{\eta}(x^{*})},
\end{equation}
for all natural parameters $\eta_{i}$, with $Q_{\eta}(x)$ the cdf and $q_{\eta}(x)$ the pdf of $x$. Similarly, it does not matter for the probit example whether we sample the $f_{i}$ jointly by sampling $x$, or whether we sample them directly and independently. After sampling the $f_{i}$, we can use $s_{i}(\eta,z_{i}^{*}) = \mu_{i} + \sigma_{i}z_{i}^{*}$ as proposed above, but we might equivalently proceed using (\ref{eq:invtrans}), or something else entirely. Finding the most efficient strategy we mostly leave for future work, although Sections \ref{hess} and \ref{lintrans} offer some further insights into what is possible.

\subsection{Using the Hessian of the log posterior}
\label{hess}

When we have both first and second order gradient information for $\log p(x,y)$ and if we choose our approximation to be multivariate Gaussian, i.e.\ $q_{\eta}(x) = N(m(\eta),V(\eta))$, we have a third option for approximating the statistics used in the regression. For Gaussian $q(x)$ and twice differentiable $\log p(x,y)$, \citet{MinkaThesis} and \citet{archambeau} show that
\begin{equation}
\nabla_{m} \E_{q}[\log p(x,y)] = \E_{q}[\nabla_{x}\log p(x,y)],
\label{op1}
\end{equation}
and
\begin{equation}
\nabla_{V} \E_{q}[\log p(x,y)] =  \frac{1}{2}\E_{q}[\nabla_{x}\nabla_{x}\log p(x,y)],
\label{op2}
\end{equation}
where $\nabla_{x}\nabla_{x}\log p(x,y)$ denotes the Hessian matrix of $\log p(x,y)$ in $x$.
\npar
For the multivariate Gaussian distribution we know that the natural parameters are given as $\eta_{1} = V^{-1}m$ and $\eta_{2} = V^{-1}$. Using this relationship, we can derive Monte Carlo estimators $\hat{g}$ and $\hat{C}$ using the identities~(\ref{eq:reggrad1}, \ref{eq:reggrad2}). We find that these stochastic approximations are often even more efficient than the ones in Section~\ref{usinggrad}, provided that the Hessian matrix of $\log p(x,y)$ can be calculated cheaply. This type of approximation is especially powerful when combined with the extension presented in the next section.

\subsection{Linear transformations of the regression problem}
\label{lintrans}
It is well known that classical linear least squares regression is invariant to invertible linear transformations of the explanatory variables. We can use the same principle in our stochastic approximation algorithm to allow us to work with alternative parameterizations of the approximate posterior $q(x)$. These alternative forms can be easier to implement or lead to more efficient algorithms, as we show in this section.
\npar
In classical linear least squares regression, we have an $N \times D$ matrix of explanatory variables $X$, and an $N \times 1$ vector of dependent variables $Y$. Instead of doing a linear regression with these variables directly, we may equivalently perform the linear regression using a transformed set of explanatory variables $\tilde{X} = XK'$, with $K$ any invertible matrix of size $D \times D$. The least squares estimator $\tilde{\beta} = (\tilde{X}'\tilde{X})^{-1}\tilde{X}'Y$ of the transformed problem can then be used to give the least squares estimator of the original problem as $\hat{\beta} = K'\tilde{\beta}$:
\[
\hat{\beta} = K'(KX'XK')^{-1}KX'Y = (KX'X)^{-1}KX'Y = (X'X)^{-1}X'Y.
\]
Using the same principle, we can rewrite the optimality condition of~\eqref{eq:vblinreg} as
\begin{equation}
\label{eq:vblinregtrans}
\tilde{\eta} = \E_{q_{\eta}}[K(\eta)\tilde{T}(x)'\tilde{T}(x)]^{-1}\E_{q_{\eta}}[K(\eta)\tilde{T}(x)'\log p(x,y)],
\end{equation}
for any invertible matrix $K$, which may depend on the variational parameters $\eta$. Instead of solving our original least squares regression problem, we may thus equivalently solve this transformed version. When we perform the linear regression in (\ref{eq:vblinregtrans}) for a fixed set of parameters $\eta$, the result will be identical to that of the original regression with $K(\eta) = \Id$, as long as we use the same random numbers for both regressions. However, when the Monte Carlo samples (`data points' in our regression) are generated using different values of $\eta$, as is the case with the proposed stochastic approximation algorithm, the two regressions will not necessarily give the same solution for a finite number of samples. If the true posterior $p(x|y)$ is of the same functional form as the approximation $q_{\eta}$, the exact convergence result of Section~\ref{sa} holds for any invertible $K(\eta)$, so it is not immediately obvious which $K(\eta)$ is best for general applications.
\npar
We hypothesize that certain choices of $K(\eta)$ may lead to statistically more efficient stochastic approximation algorithms for certain specific problems, but we will not pursue this idea here. What we will discuss is the observation that the stochastic approximation algorithm may be easier to implement for some choices of $K(\eta)$ than for others, and that the computational costs are not identical for all $K(\eta)$. In particular, the transformation $K(\eta)$ allows us to use different parameterizations of the variational approximation. Let $q_{\phi}$ be such a reparameterization of the approximation, let the new parameter vector $\phi(\eta)$ be an invertible and differentiable transformation of the original parameters $\eta$, and set $K(\eta)$ equal to the inverse transposed Jacobian of this transformation, i.e.\ $K(\eta) = [\nabla_{\eta} \phi(\eta)]^{-1}$. Using the properties of the exponential family of distributions, we can then show that
\begin{equation} \label{eq:transstat}
K(\eta)\Cov_{q_{\phi}}[T(x),h(x)] = \nabla_{\phi} \E_{q_{\phi}}[h(x)], 
\end{equation}
for any differentiable function $h(x)$. Using this result, the stochastic approximations of Section~\ref{usinggrad} for the transformed regression problem are
\begin{eqnarray} \label{eq:grad2}
x^{*} & = & s(\phi,z^{*}), \text{ with $s()$ and $z^{*}$ such that }  x^{*} \sim q_{\phi}(x) \\
\hat{g} & = & \nabla_{\phi} \log p(s(\phi,z^{*}),y) \\
\hat{C} & = &  \nabla_{\phi} T(s(\phi,z^{*})).
\end{eqnarray}
These new expressions for $\hat{g}$ and $\hat{C}$ may be easier to calculate than the original ones \eqref{eq:grad}, and the resulting $\hat{C}$ may have a structure making it easier to invert in some cases. An example of this occurs when we use a Gaussian approximation in combination with the stochastic approximations of Section~\ref{hess}, using the gradient and Hessian of $\log p(x,y)$. In this case we may work in the usual natural parameterization, but doing so gives a dense matrix $\hat{C}$ with dimensions proportional to $M^{2}$, where $M$ is the dimension of $x$. For large $M$, such a stochastic approximation is expensive to store and invert. However, using the stochastic approximations above, we may alternatively parameterize our approximation in terms of the mean $m$ and variance $V$. Working in this parameterization, we can express the update equations for the natural parameters in terms of the gradient and Hessian of $\log p(x,y)$ and the average sampled $x$ value, instead of the (higher dimensional) $g$ and $C$ statistics. The resulting algorithm, as derived in Appendix~\ref{app:gva}, is therefore more efficient in terms of both computation and storage. Pseudocode for the new algorithm is given below.

\begin{algorithm}[H]
\caption{Stochastic Approximation for Gaussian Variational Approximation}
\label{algo:gaussvb}
\begin{algorithmic}

\REQUIRE An unnormalized, twice differentiable posterior distribution $p(x,y)$
\REQUIRE The total number of iterations $N$

\STATE Initialize the mean and variance of the approximation ($m_{1}, V_{1}$) to a first guess, for example by matching the prior $p(x)$
\STATE Initialize $z_{1} = m_{1}$, $P_{1} = V_{1}^{-1}$ and $a_{1} = 0$
\STATE Initialize $\bar{z} = 0$, $\bar{P} = \mathbf{0}$ and $\bar{a} = 0$
\STATE Set step-size $w = 1/\sqrt{N}$

\FOR{$t = 1:N$}

\STATE Generate a draw $x_{t}^{*}$ from $N(m_{t},V_{t})$

\STATE Calculate the gradient $g_{t}$ and Hessian $H_{t}$ of $\log p(x,y)$ at $x_{t}^{*}$

\STATE Set $a_{t+1} = (1-w)a_{t} + wg_{t}$
\STATE Set $P_{t+1} = (1-w)P_{t} - wH_{t}$
\STATE Set $z_{t+1} = (1-w)z_{t} + wx_{t}^{*}$

\STATE Set $V_{t+1} = P_{t+1}^{-1}$ and $m_{t+1} = V_{t+1}a_{t+1} + z_{t+1}$

\IF{$t > N/2$}

\STATE Set $\bar{a} = \bar{a} + \frac{2}{N} g_{t}$
\STATE Set $\bar{P} = \bar{P} - \frac{2}{N} H_{t}$
\STATE Set $\bar{z} = \bar{z} + \frac{2}{N} x_{t}^{*}$

\ENDIF

\ENDFOR

\STATE Set $V = \bar{P}^{-1}$ and $m = V\bar{a}+\bar{z}$
\RETURN $m,V$

\end{algorithmic}
\end{algorithm}

Instead of storing and inverting the full $C$ matrix, this algorithm uses the sparsity induced by the transformation $K(\eta)$ to work with the precision matrix $P$ instead. The dimensions of this matrix are equal to the dimension of $x$, rather than its square, providing great savings. Moreover, while the $C$ matrix in the original parameterization is always dense, $P$ will have the same sparsity pattern as the Hessian of $\log p(x,y)$, which may reduce the costs of storing and inverting it even further for many applications.
\npar
Figure~\ref{fig:probitRMSE} shows the performance of Algorithm~\ref{algo:gaussvb} as applied to our probit regression example. As is typical for this version of the algorithm, it performs even better than the algorithm using only the gradient and factor structure of the posterior distribution. Since this type of approximation is also very easy to implement efficiently in a matrix programming language like MATLAB, it also runs significantly faster than the VBEM algorithm for this example. Moreover, the algorithm is now again completely general and does not make any assumptions as to the structure of the posterior distribution (other than it being twice differentiable). This means it can easily be used for Gaussian variational approximation of almost any posterior distribution.

\subsection{Subsampling the data: double stochastic approximation} \label{sec:subsample}
The stochastic approximations derived above are all linear functions of $\log p(x,y)$ and its first and second derivatives. This means that these estimates are still unbiased even if we take $\log p(x,y)$ to be a noisy unbiased estimate of the true log posterior, rather than the exact log posterior. For most statistical applications $\log p(x,y)$ itself is a separable additive function of a number of independent factors, i.e.\ $\log p(x,y) = \sum_{j=1}^{N} \log \phi_{j}(x,y)$ as explained in Section~\ref{factorstructure}. Using this fact we can construct an unbiased stochastic approximation of $\log p(x,y)$ as
\begin{equation}
\label{subsampling}
\log \tilde{p}(x,y) = \frac{N}{K} \sum_{j=1}^{K} \log \phi_{j}(x,y),
\end{equation}
where the $K$ factors $\log \phi_{j}(x,y)$ are randomly selected from the total $N$ factors. This approach was previously proposed for online learning of topic models by~\citet{hoffman2010online}. Since $\log \tilde{p}(x,y)$ has $\log p(x,y)$ as its expectation, performing stochastic approximation based on $\tilde{p}(x,y)$ converges to the same solution as when using $p(x,y)$, provided we resample the factors in $\log \tilde{p}(x,y)$ at every iteration. By subsampling the $K \ll N$ factors in the model, the individual steps of the optimization procedure become more noisy, but since we can calculate $\tilde{p}(x,y)$ faster than we can $p(x,y)$, we can perform a larger number of steps in the same amount of time. In practice this tradeoff often favors using subsampling, and this principle has been used in many successful applications of stochastic gradient descent, see e.g.\ \citet{bottou}.
\npar
For our probit regression example we implement subsampling by dividing the sample into 10 equally sized `minibatches' of data. During each iteration of the algorithm, these minibatches are processed in random order, using Algorithm~\ref{algo:gaussvb} combined with~\eqref{subsampling} to update the variational parameters after each minibatch. As can be seen in Figure~\ref{fig:probitRMSE} this approach allows us to get a good approximation to the posterior very quickly: reaching the accuracy of converged VBEM now only requires three passes over the training data, although final convergence is not much faster than when using the full sample.

\clearpage
\section{Extensions II: Using mixtures of exponential family distributions}
\label{mix}

So far, we have assumed that the approximating distribution $q_{\eta}(x)$ is a member of the exponential family. Here we will relax that assumption. If we choose a non-standard approximation, certain moments or marginals of $q_{\eta}(x)$ are typically no longer available analytically, which should be taken into account when choosing the type of approximation. However, if we can at least sample directly from $q_{\eta}(x)$, it is often still much cheaper to approximate these moments using Monte Carlo than it would be to approximate the corresponding moments of the posterior using MCMC or other indirect sampling methods. We have identified two general strategies for constructing useful non-standard posterior approximations which are discussed in the following two sections.

\subsection{Hierarchical approximations}
\label{condstand}
If we split our vector of unknown parameters $x$ into $p$ non-overlapping blocks, our approximating posterior may be decomposed as
\[
q(x) = q(x_{1})q(x_{2}|x_{1})q(x_{3}|x_{1},x_{2}) \ldots q(x_{p}|x_{1},\ldots,x_{p-1}).
\]
If we then choose every conditional posterior $q(x_{i}|x_{1},\ldots,x_{i-1})$ to be an analytically tractable member of the exponential family, we can easily sample from the joint $q(x)$, while still having much more freedom in capturing the dependence between the different blocks of $x$. In practice, such a conditionally tractable approximation can be achieved by specifying the sufficient statistics of each exponential family block $q(x_{i}|x_{1},\ldots,x_{i-1})$ to be a function of the preceding elements $x_{1},x_{2},\ldots,x_{i-1}$. This leads to a natural type of approximation for hierarchical Bayesian models, where the hierarchical structure of the prior often suggests a good hierarchical structure for the posterior approximation.
\npar
If every conditional $q(x_{i}|x_{1},\ldots,x_{i-1})$ is in the exponential family, the joint may not be if the normalizing constant of any of those conditionals is a non-separable function of the preceding elements $x_{1},x_{2},\ldots,x_{i-1}$ and the variational parameters. However, because the conditionals are still in the exponential family, our optimality condition still holds separately for the variational parameters of each conditional with only slight modification. Taking again the derivative of the KL-divergence and setting it to zero yields:
\begin{eqnarray}
\label{eq:hierarchicalupdate}
\eta_{i} & = & C_{i}^{-1}g_{i} \\
C_{i} & = & \E_{q(x_{1},\ldots,x_{i-1})}\{ \Var_{q(x_{i}|x_{1},\ldots,x_{i-1})}[T_{i}(x_{i})] \} \nonumber\\
g_{i} & = & \E_{q(x_{1},\ldots,x_{i-1})}\{ \Cov_{q(x_{i},\ldots,x_{p}|x_{1},\ldots,x_{i-1})}[T_{i}(x_{i}) , r_{-i}(x)] \}, \nonumber\\
r_{-i}(x) & = & \log p(x,y) - \log q_{\eta}(x_{1},\ldots,x_{i-1}) - \log q_{\eta}(x_{i+1},\ldots,x_{p}|x_{1},\ldots,x_{i}) \nonumber\\
& = & \log p(x,y) - \log q_{\eta}(x) + \log q_{\eta}(x_{i}|x_{1},\ldots,x_{i-1}), \nonumber
\end{eqnarray}
where $T_{i}(x_{i})$ and $\eta_{i}$ denote the sufficient statistics and corresponding natural parameters of the $i$-th conditional approximation $q(x_{i}|x_{1},\ldots,x_{i-1})$, and where $r_{-i}(x)$ can be seen as the residual of the approximation with the $i$-th block left out. Note that we cannot rewrite this expression as a linear regression any further, like we did in Section~\ref{vb}, since the intercept of such a regression is related to the normalizing constant of $q(x_{i}|x_{1},\ldots,x_{i-1})$ which may now vary in $x_{1},\ldots,x_{i-1}$. However, $C_{i}$ and $g_{i}$ can still be approximated straightforwardly using Monte Carlo, and Algorithm~\ref{algo:linregvb} can still be used with these approximations, performing separate `regressions' for all conditionals during each iteration like we proposed for factorized $p(x,y)$ in Section~\ref{factorstructure}. Alternatively, Algorithm~\ref{algo:gaussvb} or any of the extensions in Section~\ref{simul} may be used to fit the different blocks of $q_{\eta}(x)$.  
\npar
Using this type of approximation, the marginals $q(x_{i})$ will generally be mixtures of exponential family distributions, which is where the added flexibility of this method comes from. By allowing the marginals $q(x_{i})$ to be mixtures with dependency on the preceding elements of $x$, we can achieve much better approximation quality than by forcing them to be a single exponential family distribution. A similar idea was used in the context of importance sampling by \citet{hoogerheide}. A practical example of this is given below.

\subsubsection{Example: A stochastic volatility model}
\label{stochvol}

\begin{figure}[ht]
	\centering
		\includegraphics[width=0.7\textwidth]{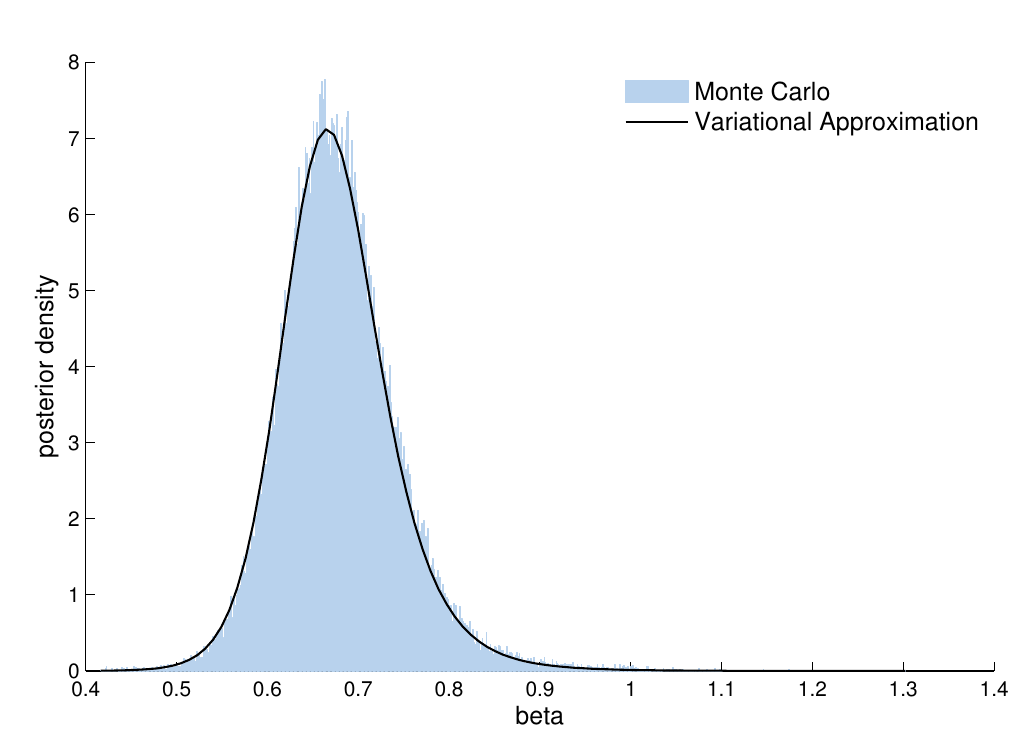}
	\caption{Exact and approximate posterior for the stochastic volatility model - $\beta$ parameter}
	\label{fig:svBeta}
\end{figure}

\begin{figure}[ht]
	\centering
		\includegraphics[width=0.7\textwidth]{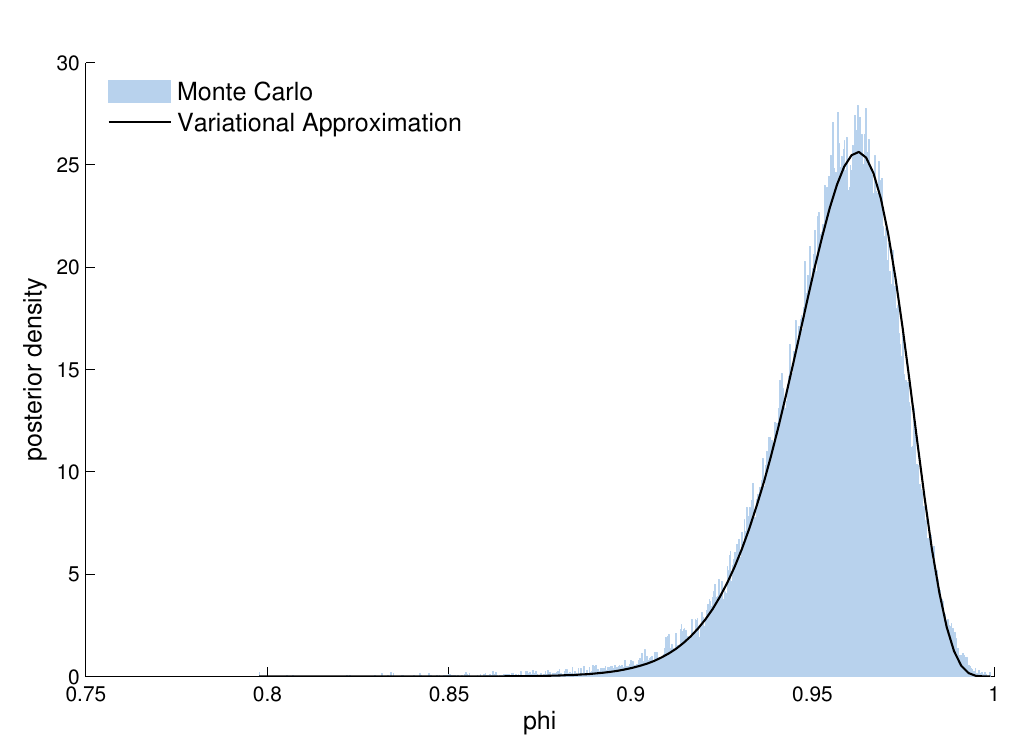}
	\caption{Exact and approximate posterior for the stochastic volatility model - $\phi$ parameter}
	\label{fig:svPhi}
\end{figure}

\begin{figure}[ht]
	\centering
		\includegraphics[width=0.7\textwidth]{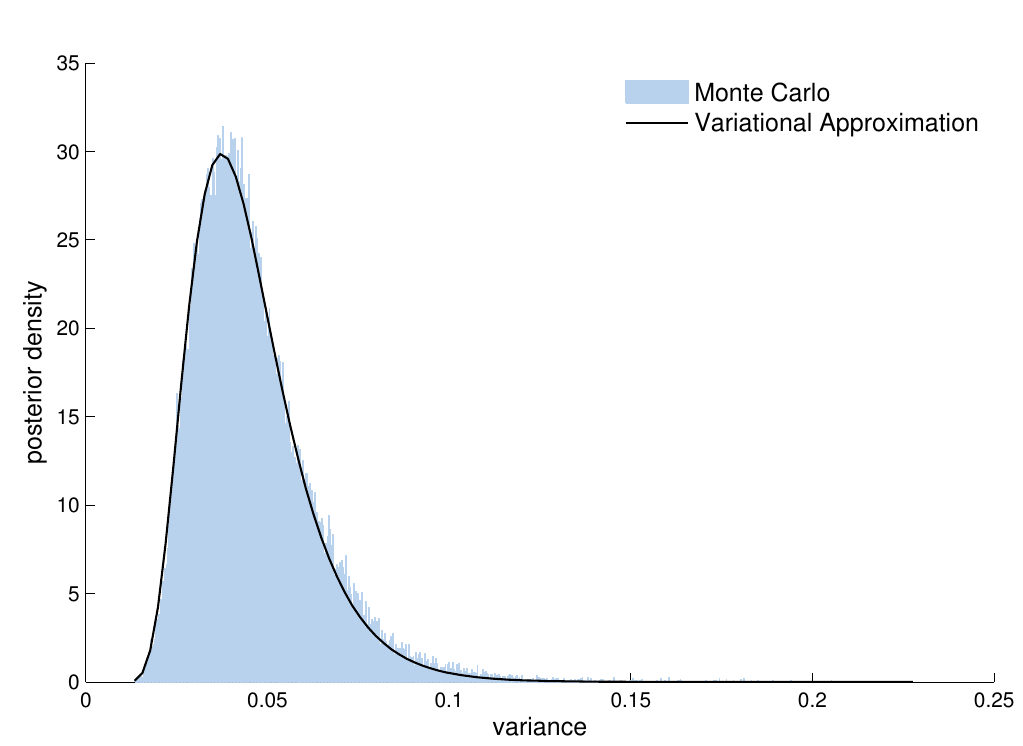}
	\caption{Exact and approximate posterior for the stochastic volatility model - $\sigma^{2}$ parameter}
	\label{fig:svVar}
\end{figure}

Stochastic volatility models for signals with time varying variances are considered extremely important in finance. Here we apply our methodology to the model and prior specified in \citet{riemannhmc}. The data we will use, from \citet{kim:98}, is the percentage change $y_{t}$ in GB Pound vs. US Dollar exchange rate, modeled as:
\[
y_{t} = \epsilon_{t}\beta\exp(v_{t}/2).
\]
The relative volatilities, $v_{t}$ are governed by the autoregressive AR(1) process
\[
v_{t+1} = \phi v_{t} + \xi_{t+1}, \text{ with } v_{1} \sim N[0,\sigma^{2}/(1-\phi^{2})].
\]
The distributions of the error terms are given by $\epsilon_{t} \sim N(0,1)$ and $\xi_{t} \sim N(0,\sigma^{2})$. The prior specification is as in \citet{riemannhmc}:
\[
p(\beta) \propto \beta^{-1}, \hspace{1cm} (\phi + 1)/2 \sim \text{Beta}(20,1.5), \hspace{1cm} \sigma^{2} \sim \text{Inv-Gamma}(5,0.25).
\]
Following the strategy outlined above, we use the hierarchical structure of the prior to suggest a hierarchical structure for the approximate posterior:
\[
q_{\eta}(\phi,\sigma^{2},\beta,v) = q_{\eta}(\phi)q_{\eta}(\sigma^{2}|\phi)q_{\eta}(\beta,v|\phi,\sigma^{2}).
\]
The prior of $\phi$ is in the exponential family, so we choose the posterior approximation $q_{\eta}(\phi)$ to be of the same form:
\[
q_{\eta}[(\phi + 1)/2] = \text{Beta}(\eta_{1},\eta_{2}).
\]
The prior for $\sigma^{2}$ is inverse-Gamma, which is also in the exponential family. We again choose the same functional form for the posterior approximation, but with a slight modification in order to capture the posterior dependency between $\phi$ and $\sigma^{2}$:
\[
q_{\eta}(\sigma^{2}|\phi) \sim \text{Inv-Gamma}(\eta_{3},\eta_{4} + \eta_{5}\phi^{2}),
\]
where the extra term $\eta_{5}\phi^{2}$ was chosen by examining the functional form of the exact full conditional $p(\sigma^{2}|\phi,v)$.
\npar
Using the notation $f = (\log(\beta),v')'$, the conditional prior $p(f|\phi,\sigma^{2})$ can be seen as the diffuse limit of a multivariate normal distribution. We therefore also use a multivariate normal conditional approximate posterior:
\[
q_{\eta}(f|\phi,\sigma^{2}) = \frac{p(f|\phi,\sigma^{2})q_{\eta}(y|f)}{q_{\eta}(y|\phi,\sigma^{2})},
\]
with $p(f|\phi,\sigma^{2})$ the Gaussian prior, $q_{\eta}(y|f)$ a Gaussian approximate likelihood of the form
\[
q_{\eta}(y|f) = (2\pi)^{-T/2}\sqrt{|\eta_{6}|}\exp\left[\eta_{7}'\eta_{6}^{-1}\eta_{7}\right]\exp\left[\eta_{7}'f -\frac{1}{2} f'\eta_{6}f\right],
\]
with $\eta_{6}$ a $T \times T$ positive-definite matrix and $\eta_{7}$ a $T \times 1$ vector, and where
\[
q_{\eta}(y|\phi,\sigma^{2}) = \int_{f} p(f|\phi,\sigma^{2})q_{\eta}(y|f) df
\]
is the normalizing constant of our posterior approximation $q_{\eta}(f|\phi,\sigma^{2})$.
\npar
Now that we have defined the functional form of the approximate posterior, we can fit its parameters by applying (\ref{eq:hierarchicalupdate}) to each of the blocks $q_{\eta}(\phi)$, $q_{\eta}(\sigma^{2}|\phi)$, and $q_{\eta}(f|\phi,\sigma^{2})$. We approximate the statistics of the first two blocks using gradients as proposed in Section~\ref{usinggrad}. The last (multivariate Gaussian) block is updated using both the gradient and the Hessian of $p(y|f)$ via the optimized expressions of Algorithm~\ref{algo:gaussvb}.
\npar
For the first block $q_{\eta}(\phi)$ this gives us the following stochastic approximations:
\begin{eqnarray}
\phi^{*} & = & s_{1}(\eta,z_{1}^{*}), \text{ with $s_{1}()$ and $z_{1}^{*}$ such that }  \phi^{*} \sim q_{\eta}(\phi) \\
\sigma^{2*} & = & s_{2}(\eta,z_{2}^{*},\phi^{*}), \text{ with $s_{2}()$ and $z_{2}^{*}$ such that }  \sigma^{2*} \sim q_{\eta}(\sigma^{2}|\phi^{*}) \\
\hat{C}_{1} & = & \nabla_{\eta}[s_{1}(\eta,z_{1}^{*})]\nabla_{\phi}[T_{1}(\phi^*)] \\
\hat{g}_{1} & = & \nabla_{\eta}[s_{1}(\eta,z_{1}^{*})]\{\nabla_{\phi}\E_{q(f|\phi^{*},\sigma^{2*})}[  \log p(\phi^{*},\sigma^{2*},f,y) - \log q_{\eta}(\sigma^{2*},f|\phi^{*})] \label{eq:sv1} \\
& & + \nabla_{\phi}[s_{2}(\eta,z_{2}^{*},\phi^{*})]\nabla_{\sigma^{2}}\E_{q(f|\phi^{*},\sigma^{2*})}[\log p(\phi^{*},\sigma^{2*},f,y) - \log q_{\eta}(\sigma^{2*},f|\phi^{*})] \} \nonumber\\
& = & \nabla_{\eta}[s_{1}(\eta,z_{1}^{*})]\{\nabla_{\phi}[\log p(\phi^{*}) + \log q_{\eta}(y|\phi^{*},\sigma^{2*}) - \log q_{\eta}(\sigma^{2*}|\phi^{*}) \label{eq:sv2} \\
& & + \E_{q(f|\phi^{*},\sigma^{2*})}(\log p(y|f) - \log q_{\eta}(y|f) ) ] \nonumber \\
& & + \nabla_{\phi}[s_{2}(\eta,z_{2}^{*},\phi^{*})]\nabla_{\sigma^{2}}[\log p(\sigma^{2*}) + \log q_{\eta}(y|\phi^{*},\sigma^{2*}) - \log q_{\eta}(\sigma^{2*}|\phi^{*}) \nonumber\\
& &  + \E_{q(f|\phi^{*},\sigma^{2*})}(\log p(y|f) - \log q_{\eta}(y|f) )]\} \nonumber\\
& \approx & \nabla_{\eta}[s_{1}(\eta,z_{1}^{*})]\{\nabla_{\phi}[\log p(\phi^{*}) + \log q_{\eta}(y|\phi^{*},\sigma^{2*}) - \log q_{\eta}(\sigma^{2*}|\phi^{*})], \label{eq:svapprox}
\end{eqnarray}
where $T_{1}(\phi^*)$ are the sufficient statistics of $q_{\eta}(\phi)$, and where we make use of the fact that
\[
p(\phi,\sigma^{2},\beta,f) = p(\phi)p(\sigma^{2})p(f|\phi,\sigma^{2})p(y|f)
\]
and
\begin{eqnarray}
q_{\eta}(\sigma^{2},f|\phi) & = & q_{\eta}(\sigma^{2}|\phi)q_{\eta}(f|\phi,\sigma^{2}) \nonumber\\
& = & q_{\eta}(\sigma^{2}|\phi)p(f|\phi,\sigma^{2})q_{\eta}(y|f)/q_{\eta}(y|\phi,\sigma^{2}). \nonumber
\end{eqnarray}
Cancelling the prior term $p(f|\phi,\sigma^{2})$ in $p()$ and $q()$ then allows us to go from \eqref{eq:sv1} to \eqref{eq:sv2}. The approximate marginal likelihood $q_{\eta}(y|\phi,\sigma^{2})$ and the expectations with respect to $q_{\eta}(f|\phi,\sigma^{2})$ can be evaluated analytically using the Kalman filter and smoother \citep[e.g.][]{koopmanbook}, which means we do not have to sample $f$ for this problem. Note that \eqref{eq:sv2} includes both the direct effect of $\phi$, as well as its indirect effects through $q_{\eta}(\sigma^{2}|\phi)$ and $q_{\eta}(f|\phi,\sigma^{2})$. If the functional form of $q()$ is close to that of $p()$, the relative importance of these indirect effects is low. In most cases we can therefore ignore these indirect effects with little to no loss of accuracy. For the current application we find that using \eqref{eq:svapprox} instead of \eqref{eq:sv2} gives virtually identical results.
\npar
The stochastic approximations for the second block $q_{\eta}(\sigma^{2}|\phi)$ are given by
\begin{eqnarray}
\hat{C}_{2} & = & \nabla_{\eta}[s_{2}(\eta,z_{2}^{*},\phi^{*})]\nabla_{\sigma^{2}}[T_{2}(\sigma^{2*})] \\
\hat{g}_{2} & = & \nabla_{\eta}[s_{2}(\eta,z_{2}^{*},\phi^{*})]\nabla_{\sigma^{2}}[\log p(\sigma^{2*}) + \log q_{\eta}(y|\phi^{*},\sigma^{2*})\nonumber\\
& & + \E_{q(f|\phi^{*},\sigma^{2*})}(\log p(y|f) - \log q_{\eta}(y|f) ) ] \nonumber\\
& \approx & \nabla_{\eta}[s_{2}(\eta,z_{2}^{*},\phi^{*})]\nabla_{\sigma^{2}}[\log p(\sigma^{2*}) + \log q_{\eta}(y|\phi^{*},\sigma^{2*})],\nonumber
\end{eqnarray}
where $T_{2}(\sigma^{2*})$ are the sufficient statistics of $q_{\eta}(\sigma^{2}|\phi)$.
\npar
Finally, the updates for the likelihood approximation (using Algorithm~\ref{algo:gaussvb}) are given by
\begin{eqnarray}
a_{t+1} & = & (1-w)a_{t} + w\E_{q_{\eta}(f|\phi^{*},\sigma^{2*})}[ \nabla_{f} \log p(y|f) ] \nonumber\\
z_{t+1} & = & (1-w)z_{t} + w\E_{q_{\eta}(f|\phi^{*},\sigma^{2*})}[ f ] \nonumber\\
\eta_{6,t+1} & = & (1-w)\eta_{6,t} - w\E_{q_{\eta}(f|\phi^{*},\sigma^{2*})}[ \nabla_{f}\nabla_{f} \log p(y|f) ] \nonumber\\
\eta_{7,t+1} & = & a_{t+1} + \eta_{6,t+1}z_{t+1}. \nonumber
\end{eqnarray}
Here again, the expectations with respect to the approximate posterior $q_{\eta}(f|\phi,\sigma^{2})$ can be calculated analytically using the Kalman filter/smoother and do not have to be approximated by sampling. Furthermore we know that the Hessian of the log likelihood is sparse, which means that only a relatively small number of the parameters in $\eta_{6}$ will be non-zero: all elements on the diagonal and all elements in the column and row belonging to $\log(\beta)$. This sparsity is also what makes fitting this posterior approximation feasible, since inverting a dense $T \times T$ precision matrix would be much too expensive. Even with this sparsity, our optimization problem is still fairly high dimensional with about 2000 free parameters. Nevertheless, we find that our approximation converges very quickly using 250 iterations of our algorithm, with a single $(\phi,\sigma^{2})$ sample per iteration, which takes our single-threaded MATLAB implementation half a second to complete on a 3GHz processor. This is more than two orders of magnitude faster than the running time required by advanced MCMC algorithms for this problem.
\npar
We compare the results of our posterior approximation against the ``true'' posterior, provided by a very long run of the MCMC algorithm of \citet{riemannhmc}. As can be seen from Figures \ref{fig:svBeta}, \ref{fig:svPhi} and \ref{fig:svVar}, the posterior approximations for the model parameters are nearly exact. Similarly, the posterior approximations for the latent volatilities $v$ (not shown) are also indistinguishable from the exact posterior.
\npar
Our approach to doing inference in the stochastic volatility model shares some characteristics with the approach of \citet{LiesenfeldEIS}. They fit a Gaussian approximation to the posterior of the volatilities for given $\phi,\sigma^{2},\beta$ parameters, using the importance sampling algorithm of \citet{RichardEIS}, which is based on auxiliary regressions somewhat similar to those in Algorithm \ref{algo:linregvb}. They then infer the model parameters using MCMC methods. The advantage of our method is that we are able to leverage the information in the gradient and Hessian of the posterior, and that our stochastic approximation algorithm allows us to fit the posterior approximation very quickly for all volatilities simultaneously, while their approach requires optimizing the approximation one volatility at a time. Unique to our approach is also the ability to concurrently fit a posterior approximation for the model parameters $\phi,\sigma^{2},\beta$ and have the approximate posterior of the volatilities depend on these parameters, while \citet{LiesenfeldEIS} need to re-construct their approximation every time a new set of model parameters is considered. As a result, our approach is significantly faster for this problem.

\FloatBarrier

\subsection{Using auxiliary variables}
\label{auxvar}
Another approach to constructing flexible posterior approximations is using the conditional exponential family approximation of Section~\ref{condstand}, but letting the first block of variables be a vector of \textit{auxiliary variables} $u$, that are not part of the original set of model parameters and latent variables, $x$. The posterior approximation then has the form
\[
q(x,u) = q(u)q(x|u).
\]
The factors $q(u)$ and $q(x|u)$ should both be analytically tractable members of the exponential family, which allows the marginal approximation $q(x)$ to be a general mixture of exponential family distributions, like a mixture of normals for example. If we use enough mixture components, the approximation $q(x)$ could then in principle be made arbitrarily close to $p(x|y)$. Note, however, that if $p(x|y)$ is multimodal our optimization problem might suffer from multiple local minima, which means that we are generally not guaranteed to find the optimal approximation. 
\npar
The mixture approximation $q(x)$ can be fitted by performing the standard KL-divergence minimization:
\begin{equation}
\label{eq:kldiv2}
\hat{\eta} = \arg\min_{\eta} \E_{q_{\eta}} [\log q_{\eta}(x) - \log p(x,y)].
\end{equation}
From (\ref{eq:kldiv2}) it becomes clear that an additional requirement of this type of approximation is that we can integrate out the auxiliary variables $u$ from the joint $q(x,u)$ in order to evaluate the marginal density $q(x)$ at a given point $x$. Fortunately this is easy to do for many interesting approximations, such as discrete mixtures of normals or continuous mixtures like Student's t distributions. Also apparent from (\ref{eq:kldiv2}) is that we cannot use this approximation directly with the stochastic approximation algorithms proposed in the last sections since $q(x)$ is itself not part of the exponential family of distributions. However, we can rewrite (\ref{eq:kldiv2}) as
\begin{equation}
\label{eq:kldiv3}
\hat{\eta} = \arg\min_{\eta} \E_{q_{\eta}} [\log q_{\eta}(x,u) - \log \tilde{p}(x,y,u)],
\end{equation}
with $\tilde{p}(x,y,u)=p(x,y)q_{\eta}(u|x)$, and
\[
q_{\eta}(u|x) = \frac{q_{\eta}(x|u)q_{\eta}(u)}{\int q_{\eta}(x|u)q_{\eta}(u) du} = \frac{q_{\eta}(x|u)q_{\eta}(u)}{q_{\eta}(x)}.
\]
Equation~\ref{eq:kldiv3} now once again has the usual form of a KL-divergence minimization where the approximation, $q_{\eta}(x,u)$, consists of exponential family blocks $q_{\eta}(u)$ and $q_{\eta}(x|u)$. By including the auxiliary variables $u$ in the `true' posterior density, we can thus once again make use of our efficient stochastic optimization algorithms. Including $u$ in the posterior does not change the marginal posterior $p(x|y)$ which is what we are interested in. We now describe a practical example of this approach using an approximation consisting of a mixture of normals.

\subsubsection{Example: A beta-binomial model for overdispersion}
\label{mixture_example}
\begin{figure}[htb]
	\centering
		\includegraphics[width=0.8\textwidth]{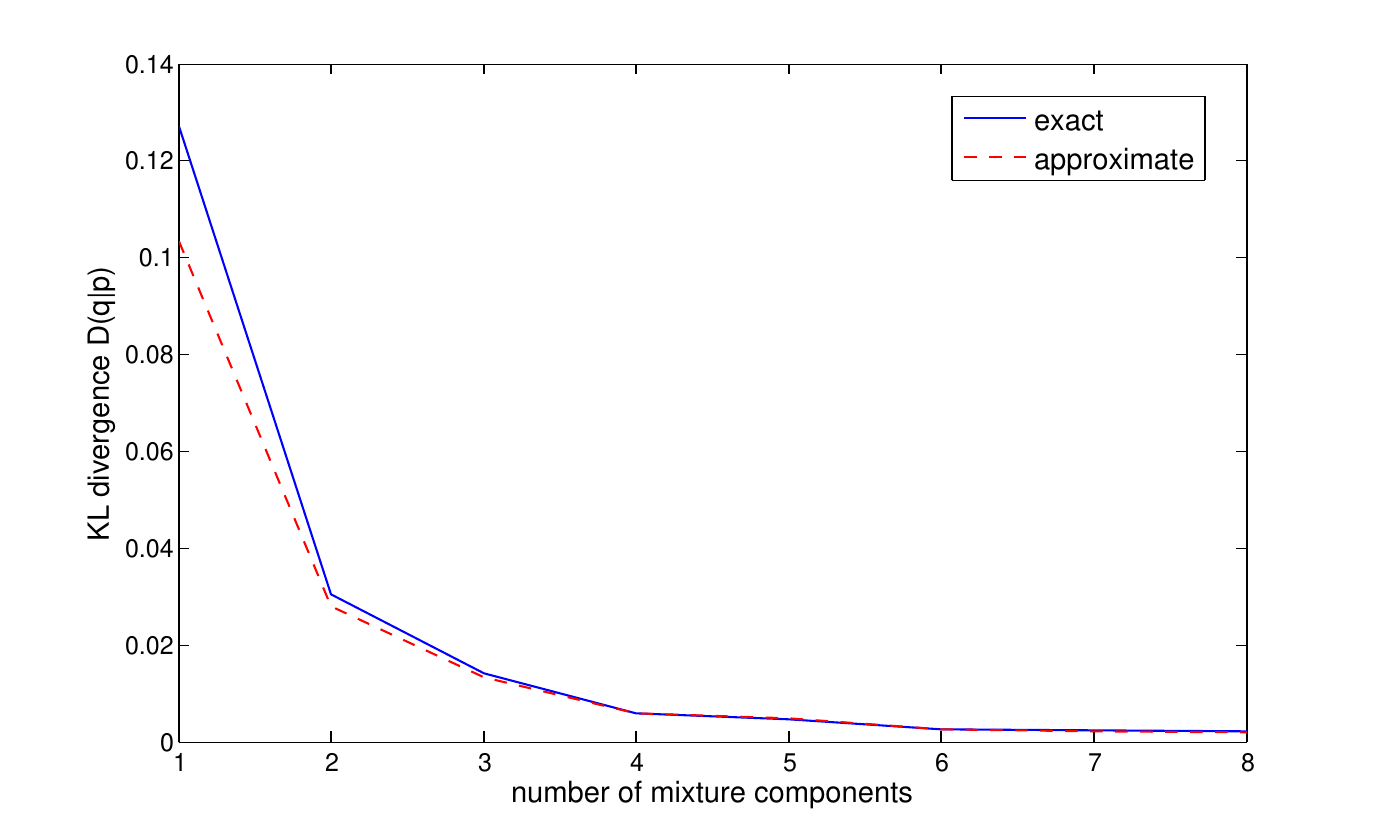}
	\caption{KL-divergence between the variational approximation and the exact posterior density for an increasing number of mixture components. The exact divergence is given by the solid blue line, while the approximation from Section~\ref{marglike} is given by the dashed red line. Note that the log marginal likelihood is given by $\log p(y) = \hat{\eta}_{0} + U(\eta) + D(q_{\eta}|p) $, with $\hat{\eta}_{0} + U(\eta) = \E_{q}[\log p(x,y) - \log q(x)]$ its usual lower bound. This means that the height of the solid blue line can also be interpreted as the approximation error of this bound for approximating the log marginal likelihood. The corresponding approximation error for the newly proposed marginal likelihood approximation (Section~\ref{marglike}, Equation~\ref{eq:newmarglikeapprox}) is then given by the difference between the solid and dashed lines: The new approximation for the marginal likelihood is thus much more accurate than the usual lower bound.}
	\label{fig:mixkldiv}
\end{figure}

\begin{figure}[htb]
	\centering
		\includegraphics[width=0.8\textwidth]{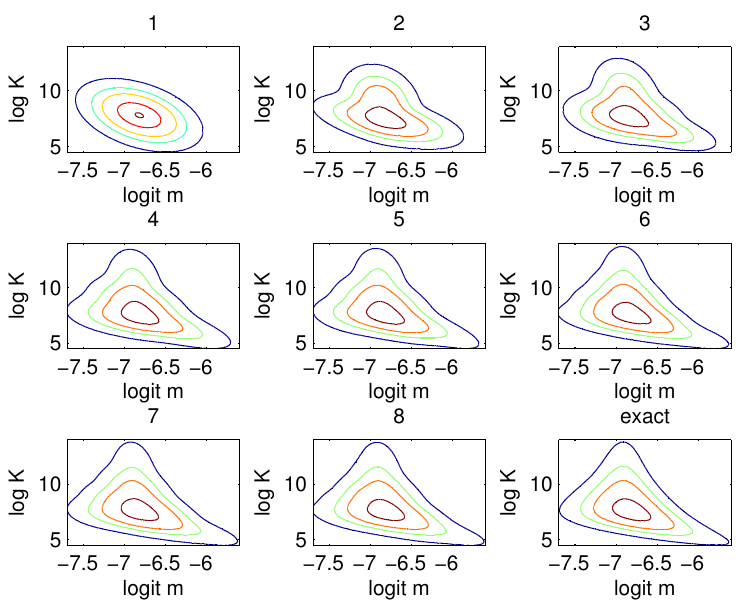}
	\caption{Contour plots of posterior approximations using 1-8 mixture components, with the exact posterior at the bottom-right. With seven or eight mixture components the approximation is visually indistinguishable from the true posterior.}
	\label{fig:mixcontours}
\end{figure}

\citet[Section 5.4]{albert:09} considers the problem of estimating the rates of death from stomach cancer for the largest cities in Missouri. This cancer mortality data is available from the R package LearnBayes, and consists of 20 pairs $(n_{j},y_{j})$ where $n_{j}$ contains the number of individuals that were at risk in city $j$, and $y_{j}$ is the number of cancer deaths that occurred in that city. The counts $y_{j}$ are overdispersed compared to what one could expect under a binomial model with constant probability, so \citet{albert:09} assumes the following beta-binomial model with mean $m$ and precision $K$:
\[ 
P(y_{j}|m,K) = \binom{n_{j}}{y_{j}}\frac{B(Km + y_{j}, K(1-m) + n_{j} - y_{j})}{B(Km, K(1-m))},
\]
where $B(\cdot,\cdot)$ denotes the Beta-function. The parameters $m$ and $K$ are given the following improper prior:
\[
p(m,K) \propto \frac{1}{m(1-m)}\frac{1}{(1+K)^{2}}.
\]
The resulting posterior distribution is non-standard and extremely skewed. To ameliorate this, \citet{albert:09} proposes the reparameterization
\[
x_{1} = \text{logit}(m), \text{ and } x_{2} = \log(K).
\]
The form of the posterior distribution $p(x|y)$ still does not resemble any standard distribution, so we will approximate it using a finite mixture of $L$ bivariate Gaussians. In order to do this, we first introduce an auxiliary variable $u$, to which we assign a categorical approximate posterior distribution with $L$ possible outcomes:
\[
q_{\eta}(u) = \exp\left[\delta(u=1)\eta_{1} + \delta(u=2)\eta_{2} + \dots + \delta(u=L)\eta_{L} - U(\eta)\right],
\]
where $\delta(.)$ is the indicator function and $U(\eta)$ is the normalizer.
\npar
Conditional on $u$, we assign $x$ a Gaussian approximate posterior
\[
q_{\eta}(x|u=i) = N(\mu_{i},\Sigma_{i}).
\]
By adapting the true posterior to include $u$ as described above, we can fit this approximate posterior to $p(x|y)$. Here, the auxiliary variable $u$ is discrete, and hence our posterior approximation is not differentiable with respect to this variable. We must therefore use the basic stochastic approximation of Section~\ref{sa} to fit $q_{\eta}(u)$. In order to reduce the variance of the resulting stochastic approximations, we Rao-Blackwellize them by taking expectations with respect to $q_{\eta}(u|x)$. If we then also take advantage of the sparsity in the covariance matrix of the sufficient statistics, this leads to the following update equations:
\begin{eqnarray}
x_{t}^{*} & \sim & q_{\eta_{t}}(x) \nonumber\\ 
\hat{C}_{t,i} & = & \E_{q_{\eta_{t}}(u|x_{t}^{*})}[ \delta(u=i) ] = q_{\eta}(u=i|x_{t}^{*}) \nonumber\\
\hat{g}_{t,i} & = & \hat{C}_{t,i}[\log p(x_{t}^{*},y) + \log q_{\eta}(u=i|x_{t}^{*}) - \log q_{\eta_{t}}(x_{t}^{*}|u=i)] \nonumber\\
 & = & \hat{C}_{t,i}[\log p(x_{t}^{*},y) + \log q_{\eta_{t}}(x_{t}^{*}|u=i) + \log q_{\eta_{t}}(u=i) \nonumber\\
& & - \log q_{\eta_{t}}(x_{t}^{*}) - \log q_{\eta_{t}}(x_{t}^{*}|u=i)] \nonumber\\
& = & \hat{C}_{t,i}[\log p(x_{t}^{*},y) - \log q_{\eta_{t}}(x_{t}^{*}) + \eta_{t,i} - U(\eta_{t})] \nonumber\\
C_{t+1,i} & = & (1-w)C_{t,i} + w\hat{C}_{t,i} \nonumber\\
g_{t+1,i} & = & (1-w)g_{t,i} + w\hat{g}_{t,i} \nonumber\\
\eta_{t+1,i} & = & \frac{g_{t+1,i}}{C_{t+1,i}}, \nonumber
\end{eqnarray}
for each mixture component $i$.
\npar
Conditional on $u$, the approximate posterior for $x$ is Gaussian, and we can therefore once again use the optimized expressions from Algorithm~\ref{algo:gaussvb} to update $q_{\eta}(x|u)$:
\begin{eqnarray}
x_{t}^{*} & \sim & q_{\eta_{t}}(x) \nonumber\\
\hat{C}_{t,i} & = & \E_{q_{\eta_{t}}(u|x_{t}^{*})}[ \delta(u=i) ] = q_{\eta_{t}}(u=i|x_{t}^{*})\nonumber\\
C_{t+1,i} & = & (1-w)C_{t,i} + w\hat{C}_{t,i} \nonumber\\
a_{t+1,i} & = & (1-w)a_{t,i} + w\hat{C}_{t,i}\nabla_{x}[\log p(x^{*},y) + \log q_{\eta_{t}}(u=i|x^{*})]  \nonumber\\
H_{t+1,i} & = & (1-w)H_{t,i} + w\hat{C}_{t,i}\nabla_{x}\nabla_{x}[\log p(x^{*},y) + \log q_{\eta_{t}}(u=i|x^{*})] \nonumber\\
z_{t+1,i} & = & (1-w)z_{t,i} + w\hat{C}_{t,i}x_{t}^{*} \nonumber\\
\Sigma_{t+1,i} & = & -C_{t+1,i}H_{t+1,i}^{-1} \nonumber\\
\mu_{t+1,i} & = & -H_{t+1,i}^{-1}a_{t+1} + \frac{z_{t+1}}{C_{t+1,i}}, \nonumber
\end{eqnarray}
for each mixture component $i$. Here we have once again Rao-Blackwellized the stochastic approximations with respect to $q_{\eta}(u|x)$, which introduced the extra variable $\hat{C}_{t,i}$ compared to Algorithm~\ref{algo:gaussvb}. Also note the presence of the $\log q_{\eta_{t}}(u=i|x^{*})$ term, which enters our equations as a result of expanding the posterior to include $u$. This term has the effect of pushing apart the different mixture components of the approximation. 
\npar
We fit the approximation $q_{\eta}(x) $ using a varying number of mixture components and examine the resulting KL-divergence to the true posterior density. Since this is a low dimensional problem, we can obtain this divergence very precisely using quadrature methods. Figures \ref{fig:mixkldiv} and \ref{fig:mixcontours} show that we can indeed approximate this skewed and heavy-tailed density very well using a large enough number of Gaussians. The R-squared of the mixture approximation with 8 components is 0.997.
\npar
Also apparent is the inadequacy of an approximation consisting of a single Gaussian for this problem, with an R-squared of only 0.82. This clearly illustrates the advantages of our approach which allows us to use much richer approximations than was previously possible. Furthermore, Figure~\ref{fig:mixkldiv} shows that the KL-divergence of the approximation to the true posterior can be approximated quite accurately using the measure developed in Section~\ref{marglike}, especially if the posterior approximation is reasonably good.
\npar
The variational optimization problem for this approximation has multiple solutions, since all Gaussian mixture components are interchangeable. Since $p(x|y)$ is unimodal, however, we find that all local optima (that we find) are equally good, and are presumably also global optima. In this case, we find that we can therefore indeed approximate $p(x|y)$ arbitrarily well by using a large enough number of mixture components.
\section{Conclusion and future work}
\label{conclusion}

We have introduced a stochastic optimization scheme for variational inference inspired by a novel interpretation of fixed-form variational approximation as linear regression of the target log density against the sufficient statistics of the approximating family. Our scheme allows very generic implementation for a wide class of models since in its most basic form only the unnormalized density of the target distribution is required, although we have shown how gradient or even Hessian information can be used if available. The generic nature of our methodology would lend itself naturally to a software package for Bayesian inference along the lines of Infer.NET~\citep{InferNET10} or WinBUGS~\citep{gilks1994language}, and would allow inference in a considerably wider range of models. Incorporating automatic differentiation in such a package could clearly be beneficial. Automatic selection of the approximating family would be very appealing from a user perspective, but could be challenging in general.
\npar
Despite its general applicability, the performance of our approach was demonstrated to be very competitive for problems where we can either decompose the posterior distribution into low dimensional factors (Section~\ref{factorstructure}), or where we can make use of the gradient and Hessian of the log posterior (Section~\ref{hess}). For those rare cases where this is not the case (e.g. high dimensional discrete distributions without factor structure) we cannot presently recommend the optimization algorithm presented in this paper. The extension of our approach to this class of problems is an important direction for future work.
\npar
We have shown it is straightforward to extend our methodology to use hierarchical structured approximations and more flexible approximating families such as mixtures. This closes the gap considerably relative to MCMC methods. Perhaps the biggest selling point of MCMC methods is that they are asymptotically exact: in practice this means simply running the MCMC chain for longer can give greater accuracy, an option not available to a researcher using variational methods. However, if we use a mixture approximating family then we can tune the computation time vs. accuracy trade off simply by varying the number of mixture components used. Another interesting direction of research along this line would be to use low rank approximating families such as factor analysis models.  
\npar
Variational inference usually requires that we use conditionally conjugate models: since our method removes this restriction several possible avenues of research are opened. For example, for MCMC methods collapsed versions of models (i.e.\ with certain parameters or latent variables integrated out) sometimes permit much more efficient inference~\citep{porteous2008fast} but adapting variational methods to work with collapsed models is complex and requires custom per model methodology~\citep{teh2007collapsed}. However, our method is indifferent to whether the model is collapsed or not, so it would be straightforward to experiment with different representations of the same model. 
\npar
It is also possible to mix our method with VBEM, for example using our method for any non-conjugate parts of the model and VBEM for variables that happen to be conditionally conjugate. This is closely related to the non-conjugate variational message passing (NCVMP) algorithm of \cite{ncvmp} implemented in Infer.NET, which aims to fit non-conjugate models while maintaining the convenient message passing formalism. NCVMP only specifies how to perform the variational optimization, not how to approximate required integrals: in Infer.NET where analytic expectations are not available quadrature or secondary variational bounds are used, unlike the Monte Carlo approach proposed here. It is still an open question how these different methods could best be combined into a joint framework.

%\clearpage

\section*{Acknowledgements}
Tim Salimans wishes to acknowledge his advisors Richard Paap and Dennis Fok, as well as the anonymous referees, for their substantial help in improving the paper. He thanks The Netherlands Organization for Scientific Research (NWO) for financially supporting this project. DAK thanks Wolfson College, Cambridge, Microsoft Research Cambridge, and the Stanford Univeristy Center for Cancer Systems Biology for funding. 

\clearpage
\appendix
\section{Unnormalized to normalized optimality condition} \label{app:normalised}

The unnormalized optimality condition in \eqref{eq:vblinreg_int} is 
\begin{align} 
\tilde{\eta} = \left[ \int \tilde{q}_{\tilde{\eta}}(x) \tilde{T}(x)'\tilde{T}(x) dx \right]^{-1}\left[\int \tilde{q}_{\tilde{\eta}}(x) \tilde{T}(x)'\log p(x,y) dx\right].
\end{align}
Clearly we can replace $\tilde{q}(x)$ by its normalized version $q(x)=\tilde{q}(x)/\exp[U(\eta)]$ since the normalizing terms will cancel. Recalling $\tilde{T}(x) = (1, T(x))$ and $\tilde{\eta} = (\eta_{0}, \eta')'$ we then have
\begin{align} 
\left[ \begin{array}{cc} 
        1 & \E[T] \\
	\E[T'] & \E[T'T] 
       \end{array} \right]^{-1}
\left( \begin{array}{c}
        \E[Y] \\
	\E[T'Y]
       \end{array}
\right)
&= 
\left( \begin{array}{c}
        \eta_0 \\
	\eta
       \end{array}
\right),
\end{align}
where $Y := \log p(x,y)$. Rearranging gives
\begin{align} 
\left( \begin{array}{c}
        \E[Y] \\
	\E[T'Y]
       \end{array}
\right)
&= 
\left[ \begin{array}{cc} 
        1 & \E[T] \\
	\E[T'] & \E[T'T] 
       \end{array} \right]
\left( \begin{array}{c}
        \eta_0 \\
	\eta
       \end{array}
\right).
\end{align}
Solving for $\eta_0$ easily gives
\begin{align}
 \eta_0 &= \E[Y] - \E[T] \eta = \E[\log p(x,y) - \log q(x)] - U(\eta)\\
 \eta &= \left( \E[T'T] - \E[T'] \E[T] \right)^{-1} ( \E[T'Y]- \E[T'] \E[Y] )\\
&= \cov(T,T)^{-1} \cov(T,Y). \label{eq:solution_normalized}
\end{align}

Note that \eqref{eq:solution_normalized}, combined with $\cov[T(x), \log q_{\eta}(x))] = \cov[T(x), T(x)]\eta$ also implies that $\cov[T(x), \log p(x,y) - \log q_{\eta}(x)] = 0$ at a solution of the KL-divergence minimization. This is the same fixed point condition used in other applications of stochastic approximation variational Bayes such as \citet{paisleyvb}.   

\section{Derivation of Gaussian variational approximation}
\label{app:gva}

For notational simplicity we will derive our stochastic approximation algorithm for Gaussian variational approximation (Algorithm~\ref{algo:gaussvb}) under the assumption that $x$ is univariate. The extension to multivariate $x$ is conceptually straightforward but much more tedious in terms of notation.
\npar
Let $p(x,y)$ be the unnormalized posterior distribution of a univariate random variable $x$, and let $q(x) = N(m,V)$ be its Gaussian approximation with sufficient statistics, $T(x) = (x, -0.5x^{2})$. In order to find the mean $m$ and variance $V$ that minimize the KL-divergence between $q(x)$ and $p(x|y)$ we solve the transformed regression problem defined in \eqref{eq:vblinregtrans}, i.e.\
\begin{align*}
\eta &= \left[ K(\eta) \Cov_{q_{\eta}}(T(x),T(x)) \right]^{-1} \left[ K(\eta) \Cov_{q_{\eta}}(T(x),\log p(x,y)) \right] \\
&= C^{-1}g
\end{align*}
where
\[
K(\eta) = [\nabla_{\eta} \phi(\eta)]^{-1},
\]
with $\phi = (\phi_1, \phi_2) = (m,V)$ the usual mean-variance parameterization and where the natural parameters are given by $\eta = (V^{-1}m, V^{-1})$. Recall identity \eqref{op1} which states that
\[
\nabla_{\phi_{1}} \E_{q_{\phi}}[h(x)] = \E_{q_{\phi}}[\nabla_{x}  h(x)],
\]
with $\phi_{1} = m$ the first element of the parameter vector $\phi$, and $g(x)$ any differentiable function. Similarly, identity \eqref{op2} reads
\[
\nabla_{\phi_{2}} \E_{q_{\phi}}[h(x)] = -\frac{1}{2}\E_{q_{\phi}}[\nabla_{x}\nabla_{x}  h(x)],
\]
with $\phi_{2} = V$ the second element of the parameter vector. Using these identities we find that the regression statistics for this optimization problem are given by
\begin{align*}
C &:= K(\eta)\Cov_{q_{\phi}}[T(x),T(x)]
= \nabla_{\phi} \E_{q_{\phi}}[T(x)]  \\
&= \E_{q_{\phi}}[\nabla_x T(x)]
= \E_{q_{\phi}} \left[ \begin{matrix} 
1 & -x \\
0 & \frac12 \end{matrix} \right] 
=  \left[ \begin{matrix} 
1 & -\E_{q_{\phi}}[x] \\
0 & \frac12 \end{matrix} \right] ,
\end{align*}
and 
\begin{align*}
%\label{eq:stat2}
g &:= K(\eta)\Cov_{q_{\phi}}[T(x),\log p(x,y)] \\
&= \nabla_{\phi} \E_{q_{\phi}}[\log p(x,y)] \\
\Rightarrow  \left[ \begin{matrix} g_1 \\ g_2 \end{matrix} \right] &= \left[ \begin{matrix} \E_q[\nabla_{x}  \log p(x,y)] \\ -\frac{1}{2}\E_q[\nabla_{x}\nabla_{x} \log p(x,y)] \end{matrix} \right].
\end{align*}
Now since $\eta = C^{-1} g$ we have
\begin{align*}
\left[ \begin{matrix} Pm \\ P \end{matrix} \right] := \left[ \begin{matrix} \eta_1 \\ \eta_2 \end{matrix} \right] 
=  \left[ \begin{matrix} 
1 & -\E_{q_{\phi}}[x] \\
0 & \frac12 \end{matrix} \right] ^{-1} \left[ \begin{matrix} g_1 \\ g_2 \end{matrix} \right] \\
\Rightarrow
\eta_2 = P = 2 g_2 = -\E_q[\nabla_{x}\nabla_{x} \log p(x,y)] \\
\eta_1 = Pm = g_1 + P^{-1} \E_q[x] = \E_q[\nabla_{x} \log p(x,y)] + P^{-1} \E_q[x] 
\end{align*} 
where $Pm$ and $P=V^{-1}$ are the natural parameters (mean times precision and precision) of the approximation. Thus the quantities we need to stochastically approximate are
\begin{align*}
a &:= \E_q[\nabla_{x} \log p(x,y)] \\
H &:= \E_q[\nabla_{x}\nabla_{x} \log p(x,y)] \\
z &:= \E_q[x] 
\end{align*}

so we have $P=-H$ and $m=P^{-1}a + z$. 

\section{Connection to Efficient Importance Sampling}
\label{eisvb}
It is worth pointing out the connection between fixed-form variational Bayes and \cites{RichardEIS} \textit{Efficient Importance Sampling} (EIS) algorithm. Although these authors take a different perspective (that of importance sampling) their goal of approximating the intractable posterior distribution with a more convenient distribution is shared with variational Bayes. Specifically, \citet{RichardEIS} choose their posterior approximation to minimize the variance of the log-weights of the resulting importance sampler. This leads to an optimization problem obeying a similar fixed-point condition as in~\eqref{eq:vblinreg}, but with the expectation taken over $p(x|y)$ instead of $q(x)$. Since sampling from $p(x|y)$ directly is not possible, they evaluate this expectation by sampling from $q(x)$ and weighting the samples using importance sampling. In practice however, these `weights' are often kept fixed to one during the optimization process in order to improve the stability of the algorithm. When all weights are fixed to one, \cites{RichardEIS} fixed-point condition becomes identical to that of~\eqref{eq:vblinreg} and the algorithm is in fact fitting a variational posterior approximation.
\npar
The connection between EIS and variational Bayes seems to have gone unnoticed until now, but it has some important consequences. It is for example well known \citep[e.g.][]{epdivergence,Nickisch2008,Turner} that the tails of variational posterior approximations tend to be thinner than those of the actual posterior unless the approximation is extremely close, which means that using EIS with the importance-weights fixed to one is not to be recommended for general applications: In the case that the posterior approximation is nearly exact, one might as well use it directly instead of using it to form another approximation using importance sampling. In cases where the approximation is not very close, the resulting importance sampling algorithm is likely to suffer from infinite variance problems. The literature on variational Bayes offers some help with these problems. Specifically, \citet{variationalmcmc} propose a number of ways in which variational approximations can be combined with Monte Carlo methods, while guarding for the aforementioned problems.
\npar
Much of the recent literature \citep[e.g.][]{teh2007collapsed,honkela} has focused on the computational and algorithmic aspects of fitting variational posterior approximations, and this work might also be useful in the context of importance sampling. Algorithmically, the `sequential EIS' approach of \citet{RichardEIS} is most similar to the non-conjugate VMP algorithm of \citet{ncvmp}. As these authors discuss, such an algorithm is not guaranteed to converge, and they present some tricks that might be used to improve convergence in some difficult cases.
\npar
The algorithm presented in this paper for fitting variational approximations is provably convergent, as discussed in Section~\ref{sa}. Furthermore, Sections~\ref{marglike} and \ref{simul} present multiple new strategies for variance reduction and computational speed-up that might also be useful for importance sampling. In this paper we will not pursue the application of importance sampling any further, but exploring these connections more fully is a promising direction for future work.

\section{Choosing an estimator} \label{sec:choosing}

\begin{figure}[H]
	\centering
		\includegraphics[width=0.60\textwidth]{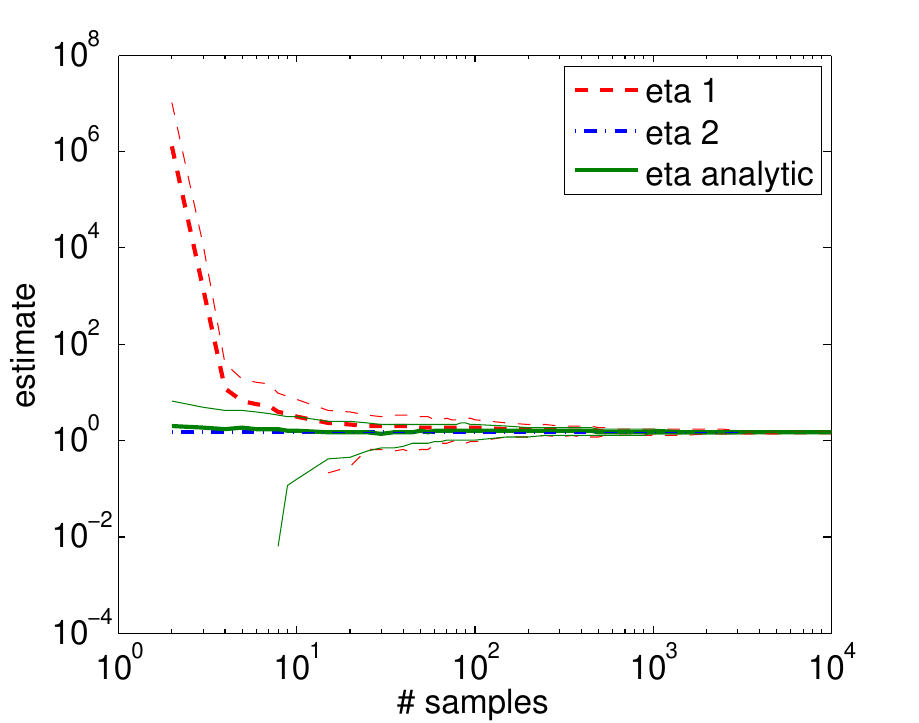}
	\caption{Comparison of three estimators for fitting a variational posterior $q$ to a simple exponential distribution $p$. $50$ repeats were used to estimate the mean and variance of the estimator: the thick line shows the mean and the thin lines show $\pm$ one standard deviation. The $x$-axis indicates the number of MC samples, $S$, used. As expected in this case $\hat{\eta}_2$ gives the correct solution of $1.5$ using $S \geq 2$ samples.}
	\label{fig:fit_exp_toy}
\end{figure}

As discussed in Section~\ref{sa}, the particular estimator used in our stochastic approximation is not the most obvious choice, but it seems to provide a lower variance approximation than other choices. In this section we consider three different MC estimators for approximating \eqref{eq:vblinreg} to see why this might be the case.
\npar
The first separately approximates the two integrals and then calculates the ratio:
\begin{align} \label{eq:est1}
 \hat{\eta}_1 &= \left( \frac1S \sum_r \tilde{T}(x_r)'\tilde{T}(x_r) \right)^{-1} \frac1S \sum_s \tilde{T}(x_s)'\log p(x_s,y) , & x_r, x_s \sim_{iid} q(x),
\end{align}
with $S$ the number of Monte Carlo samples. The second approximates both integrals using the same samples from $q$:
\begin{align} \label{eq:est2}
 \hat{\eta}_2 &= \left( \frac1S \sum_s \tilde{T}(x_s)'\tilde{T}(x_s) \right)^{-1} \frac1S \sum_s \tilde{T}(x_s)'\log p(x_s,y), & x_s \sim_{iid} q(x). 
\end{align}
Only this estimator is directly analogous to the linear regression estimator. The third estimator is available only when the first expectation is available analytically:
\begin{align} \label{eq:esta}
 \hat{\eta}_a &= \E _q\left[\tilde{T}(x)'\tilde{T}(x)\right]^{-1} \frac1S \sum_s \tilde{T}(x_s)'\log p(x_s,y), & x_s \sim_{iid} q(x).
\end{align}
We wish to understand the bias/variance tradeoff inherent in each of these estimators. To keep notation manageable consider the case with only $k=1$ sufficient statistic\footnote{These results extend in a straightforward manner to the case where $k>1$.} and let 
\begin{align}
 a(x) &= \tilde{T}(x)'\tilde{T}(x) = \tilde{T}(x)^2 \\
 b(x) &= \tilde{T}(x)  \log p(x,y).
\end{align}
We can now write the three estimators of $\eta$ more concisely as 
\begin{align}
 \hat{\eta}_1 = \frac{ \frac1S \sum_r b(x_r) }{ \frac1S \sum_s a(x_s) }, && x_r, x_s \sim_{iid} q(x) \\
 \hat{\eta}_2 = \frac{ \frac1S \sum_s b(x_s) }{ \frac1S \sum_s a(x_s) }, && x_s \sim_{iid} q(x) \\
 \hat{\eta}_a = \frac{ \frac1S \sum_s b(x_s) }{  \E[a] }, && x_s \sim_{iid} q(x).
\end{align}
Using a simple Taylor series argument it is straightforward to approximate the bias and variance of these estimators. We first consider the bias. Consider the multivariate Taylor expansion of $f : \mathbb{R}^K \rightarrow \mathbb{R}$ around the point $\bar{y} \in \mathbb{R}^K $:
\begin{align}
 f(y) \approx f(\bar{y}) + (y-\bar{y})' f'(\bar{y}) + \frac12  \text{tr}( (y-\bar{y}) (y-\bar{y})'  \nabla^2 f(\bar{y}) ).
\end{align}
From this we can derive expressions for the expectation of $f(y)$:
\begin{align}
 \E[f] & \approx f(\bar{y}) + \frac12 \text{tr}( \cov(y) f''(\bar{y}))
%  \E f \approx f(\E y) + \frac12 \sum_{ij} \cov(y)_{ij} \nabla^2 f(m)_{ij} % clearer? 
\end{align}
where we have chosen to perform the Taylor expansion around the mean $\bar{y}=\E[y]$. For the first estimator let $y = \frac1S \sum_s a(x_s)$ and $f(y)=1/y$, then we find 
\begin{align}
  \E[\hat{\eta}_1] &= \E\left[ \left( \frac1S \sum_s a(x_s) \right)^{-1} \right] \E[b] \\
&\approx \left( \frac1{\E[a]} + \frac{\var(a) }{ S \E[a]^3} \right) \E[b] \\
&= \eta + \frac{\var(a) \E[b] }{S \E[a]^3}
\end{align}
since $\var(y) = \var(a)/S$. We see that the bias term depends on the ratio $\var(a) / \E[a]^2$, i.e.\ the spread of the distribution of $a$ relative to its magnitude. 
\npar
Now for the second estimator let
\begin{align} \label{eq:y_for_eta2}
 y= \left[ \begin{array}{c} \frac1S \sum_s a(x_s) \\ \frac1S \sum_s b(x_s) \end{array} \right] 
\end{align}
so that $\eta_2 = f(y) = \frac{y_2}{y_1}$. Note that $ \cov(y) = \frac1S \cov([a,b]')$ and 
\begin{align}
 \nabla^2 f(y) = \left[ \begin{array}{cc} \frac{2y_2}{y_1^3} & -\frac1{y_1^2} \\ 
			 -\frac1{y_1^2}  & 0 
                        \end{array} \right].
\end{align}
Putting everything together we have
\begin{align}
 \E[\hat{\eta}_2] \approx  \eta + \frac{\var(a) \E b }{ S \E[a]^3} - \frac{ \cov(a,b)}{S \E[a]^2 }.
\end{align}
Note that we recover the expression for $ \E\hat{\eta}_1$ if $\cov(a,b)=0$, which makes sense because if we use different randomness for calculating $\E[a]$ and $\E[b]$ then $a,b$ have $0$ covariance in our MC estimate. Finally the analytic estimator is unbiased:
\begin{align}
  \E\hat{\eta}_a = \eta.
\end{align}
We now turn to the variances. The analytic estimator is a standard MC estimator with variance
\begin{align}
 \var(\hat{\eta}_a) & = \frac{\var(b)}{S\E[a]^2}.
\end{align}
Consider only the linear terms of the Taylor expansion:
\begin{align}
  f(y) \approx f(\bar{y}) + (y-\bar{y})' f'(\bar{y}).
\end{align}
Substituting this into the formula for variance gives
\begin{align} \label{eq:taylor-variance}
 \var[f(y)] &= \E[ (f(y)-\E[f(y)])(f(y)-\E[f(y)])' ] \\
&\approx \E[ f'(\bar{y})' (y-\bar{y})(y-\bar{y})'  f'(\bar{y}) ]   \\
&= f'(\bar{y})' \var(y) f'(\bar{y}).
\end{align}
We will calculate the variance of the second estimator and derive the variance of the first estimator from this. Again let $y$ be as in \eqref{eq:y_for_eta2}. Note that $\var(y) = \cov(a,b)/S$. We find
\begin{align} \label{eq:variance-eta2}
\var{\hat{\eta}_2} &\approx \frac1S \left( \frac{\E[b]^2 \var a}{\E[a]^4} - 2 \frac{\E[b] \cov(a,b)}{\E[a]^3} + \frac{\var b}{\E[a]^2} \right). 
\end{align}
The final term is equal to that for the analytic estimator. The second term is not present in the variance of the first estimator, since then $a$ and $b$ have no covariance under the sampling distribution, i.e.\ 
\begin{align}
\var{\hat{\eta}_1} &\approx \frac1S \left( \frac{\E[b]^2 \var a}{\E[a]^4} + \frac{\var b}{\E[a]^2} \right). 
\end{align}
The first term is always positive, suggesting that $\hat{\eta}_1$ is dominated by the analytic estimator.
\npar
Summarizing these derivations, we have
\begin{eqnarray}
\label{eq:bias}
 \text{bias}(\hat{\eta}_1) & \approx & \frac{\var(a) \E[b] }{ S \E[a]^3} \nonumber \\
 \text{bias}(\hat{\eta}_2) & \approx & \frac{\var(a) \E[b] }{ S \E[a]^3} - \frac{ \cov(a,b)}{S \E[a]^2 }.
\end{eqnarray}
Note that the first term is shared, but the first estimator does not have the covariance term as a result of the independent sampling in approximating the numerator and denominator. In contrast $\hat{\eta}_a$ is unbiased. Now consider the variances
\begin{align}
\var(\hat{\eta}_1) &\approx \frac1S \left( \frac{\E[b]^2 \var(a)}{\E[a]^4} + \frac{\var(b)}{\E[a]^2} \right)  \\
\var(\hat{\eta}_2) &\approx \frac1S \left( \frac{\E[b]^2 \var(a)}{\E[a]^4} - 2 \frac{\E[b] \cov(a,b)}{\E[a]^3} + \frac{\var(b)}{\E[a]^2} \right) \\
\var(\hat{\eta}_a) & = \frac{\var(b)}{S\E[a]^2}.
\end{align}
All three estimators have the same final term (the variance of the ``analytic'' estimator). Again the second estimator has an additional term resulting from the covariance between $a$ and $b$ which we find is typically beneficial in that it results in the variance of $\hat{\eta}$ being significantly smaller. It is worth recalling that the mean squared error (MSE) of an estimator is given by
\begin{align}
 \E[ (\eta - \hat{\eta})^2 ] &= \var(\hat{\eta}) + \text{bias}(\hat{\eta})^2.
\end{align}
Since both the variance and bias are $O(1/S)$, the variance contribution to the MSE is $O(1/S)$ whereas the bias contribution is $O(1/S^2)$, so the variance is actually a greater problem than the bias. From these expressions it is still not immediately obvious which estimator we should use. However, consider the case when the target distribution $p$ is in the same exponential family as $q$, i.e.\ when $\log p(x,y) = \tilde{T}(x)\lambda$. It is then straightforward to show that
\begin{align}
 \text{bias}(\hat{\eta}_1)&\approx\frac{\lambda \var(\tilde{T}^2)}{ S \E[\tilde{T}^2]^2} ,&\var(\hat{\eta}_1) \approx  2 \frac{ \lambda^2 \var(\tilde{T}^2) }{ S \E [ \tilde{T}^2 ]^2 } \\
 \text{bias}(\hat{\eta}_2)&\approx 0  ,& \var(\hat{\eta}_2)\approx   0 \label{eq:eta2varbias} \\
 \text{bias}(\hat{\eta}_a)&= 0 ,&\var(\hat{\eta}_a)  =  \frac{ \lambda^2 \var(\tilde{T}^2) }{ S \E [ \tilde{T}^2 ]^2 }.
\end{align}
We see that in this case for $\hat{\eta}_2$ the positive and negative contributions to both the bias and variance cancel. While this result will not hold exactly for cases of interest, it suggests that for exponential families which are capable of approximating $p$ reasonably well, $\hat{\eta}_2$ should perform significantly better than $\hat{\eta}_1$ or even $\hat{\eta}_a$. If $q$ and $p$ are of the same exponential family, it is actually possible to see that $\hat{\eta}_2$ will in fact give the exact solution in $k+1$ samples (with $k$ the number of sufficient statistics), while the other estimators have non-vanishing variance for a finite number of samples. This means that the approximate equality in \eqref{eq:eta2varbias} can be replaced by exact equality. Using $k+1$ samples $x_{i}, i=1,...,k+1$, assumed to be unique (which holds almost surely for continuous distributions $q$), we have
\begin{align} \label{eq:exact_linreg}
  \hat{\eta}_2 = \left(\sum_{i=1}^{k+1} \tilde{T}(x_{i})'\tilde{T}(x_{i})\right)^{-1} \sum_{i=1}^{k+1} \tilde{T}(x_{i})'\tilde{T}(x_{i}) \lambda =  \lambda.
\end{align}
That is, the algorithm has recovered $p(x,y)$ exactly with probability one. If we assume we know how to normalize $q$, this means we also have $p(x|y)$ exactly in this case. Note that we recover the exact answer here because the $p(x,y)$ function evaluations are in themselves \emph{noise free}, so the regression analogy really corresponds to a noise free regression.
\npar
We test the three estimators in \eqref{eq:est1}, \eqref{eq:est2} and \eqref{eq:esta} on the trivial exponential example of Section~\ref{sa} when the true exponential rate is $\lambda=1.5$, and sampling from the optimal $q$ distribution with $\eta=1.5$. The results confirm that $\hat{\eta}_2$ finds the exact rate using just $S=2$ MC samples, as predicted by \eqref{eq:exact_linreg}. We would expect $\hat{\eta}_a$ to be unbiased, and this is borne out by the results shown in Figure~\ref{fig:fit_exp_toy}. The estimator $\hat{\eta}_1$ has both poor bias and such large variance that it often gives an invalid negative rate if fewer than 10 MC samples are used. While this is clearly a very simple example it hopefully emphasizes the potential benefit to be gained from using estimators related to $\hat{\eta}_2$. 

\clearpage
\bibliographystyle{ba}
\bibliography{biball}

\begin{thebibliography}{40}
\newcommand{\enquote}[1]{``#1''}
\expandafter\ifx\csname natexlab\endcsname\relax\def\natexlab#1{#1}\fi
\expandafter\ifx\csname url\endcsname\relax
  \def\url#1{{\tt #1}}\fi
\expandafter\ifx\csname urlprefix\endcsname\relax\def\urlprefix{URL }\fi

\bibitem[{Albert(2009)}]{albert:09}
Albert, J. (2009).
\newblock {\em Bayesian Computation with R\/}.
\newblock Springer Science, New York. Second edition.

\bibitem[{Amari(1997)}]{amaring}
Amari, S. (1997).
\newblock \enquote{Neural Learning in Structured Parameter Spaces - Natural
  {R}iemannian Gradient.}
\newblock In {\em Advances in Neural Information Processing Systems\/},
  127--133. MIT Press.

\bibitem[{Attias(2000)}]{Attias2000}
Attias, H. (2000).
\newblock \enquote{A variational {B}ayesian framework for graphical models.}
\newblock In {\em Advances in Neural Information Processing Systems (NIPS)
  12\/}, 209--215.

\bibitem[{Beal and Ghahramani(2002)}]{beal2003vbem}
Beal, M.~J. and Ghahramani, Z. (2002).
\newblock \enquote{The variational {B}ayesian {EM} algorithm for incomplete
  data: with application to scoring graphical model structures.}
\newblock In {\em Bayesian Statistics 7: Proceedings of the 7th Valencia
  International Meeting\/}, 453--463.

\bibitem[{Beal and Ghahramani(2006)}]{Beal2006}
--- (2006).
\newblock \enquote{Variational {B}ayesian learning of directed graphical models
  with hidden variables.}
\newblock {\em {B}ayesian Analysis\/}, 1(4): 793--832.

\bibitem[{Bishop(2006)}]{bishop2006pattern}
Bishop, C.~M. (2006).
\newblock {\em Pattern recognition and machine learning\/}, volume~1.
\newblock Springer New York.

\bibitem[{Bottou(2010)}]{bottou}
Bottou, L. (2010).
\newblock \enquote{Large-Scale Machine Learning with Stochastic Gradient
  Descent.}
\newblock In {\em Proceedings of the 19th International Conference on
  Computational Statistics (COMPSTAT'2010)\/}, 177--187. Springer.

\bibitem[{de~Freitas et~al.(2001)de~Freitas, H{\o}jen-S{\o}rensen, Jordan, and
  Russell}]{variationalmcmc}
de~Freitas, N., H{\o}jen-S{\o}rensen, P., Jordan, M.~I., and Russell, S.
  (2001).
\newblock \enquote{Variational MCMC.}
\newblock In {\em Proceedings of the Seventeenth Conference on Uncertainty in
  Artificial Intelligence\/}, UAI'01, 120--127. San Francisco, CA, USA: Morgan
  Kaufmann Publishers Inc.

\bibitem[{Durbin and Koopman(2001)}]{koopmanbook}
Durbin, J. and Koopman, S. (2001).
\newblock {\em {T}ime {S}eries {A}nalysis by {S}tate {S}pace {M}ethods\/}.
\newblock Oxford University Press.

\bibitem[{Geweke(2005)}]{gewekebook}
Geweke, J. (2005).
\newblock {\em Contemporary {Bayesian} Econometrics and Statistics\/}.
\newblock Wiley-Interscience.

\bibitem[{Gilks et~al.(1994)Gilks, Thomas, and
  Spiegelhalter}]{gilks1994language}
Gilks, W., Thomas, A., and Spiegelhalter, D. (1994).
\newblock \enquote{A language and program for complex Bayesian modelling.}
\newblock {\em The Statistician\/}, 169--177.

\bibitem[{Girolami and Calderhead(2011)}]{riemannhmc}
Girolami, M. and Calderhead, B. (2011).
\newblock \enquote{Riemann manifold {L}angevin and {H}amiltonian {M}onte
  {C}arlo methods.}
\newblock {\em Journal of the Royal Statistical Society: Series B (Statistical
  Methodology)\/}, 73(2): 123--214.

\bibitem[{Hoffman et~al.(2010)Hoffman, Blei, and Bach}]{hoffman2010online}
Hoffman, M., Blei, D., and Bach, F. (2010).
\newblock \enquote{Online learning for latent {D}irichlet allocation.}
\newblock {\em Advances in Neural Information Processing Systems\/}, 23.

\bibitem[{Hoffman et~al.(2012)Hoffman, Blei, Wang, and
  Paisley}]{hoffman2012stochastic}
Hoffman, M., Blei, D., Wang, C., and Paisley, J. (2012).
\newblock \enquote{Stochastic Variational Inference.}
\newblock {\em arXiv preprint arXiv:1206.7051\/}.

\bibitem[{Honkela et~al.(2010)Honkela, Raiko, Kuusela, Tornio, and
  Karhunen}]{honkela}
Honkela, A., Raiko, T., Kuusela, M., Tornio, M., and Karhunen, J. (2010).
\newblock \enquote{Approximate {R}iemannian Conjugate Gradient Learning for
  Fixed-Form Variational Bayes.}
\newblock {\em Journal of Machine Learning Research\/}, 3235--3268.

\bibitem[{Hoogerheide et~al.(2012)Hoogerheide, Opschoor, and van
  Dijk}]{hoogerheide}
Hoogerheide, L., Opschoor, A., and van Dijk, H.~K. (2012).
\newblock \enquote{A class of adaptive importance sampling weighted \{EM\}
  algorithms for efficient and robust posterior and predictive simulation.}
\newblock {\em Journal of Econometrics\/}, 171(2): 101 -- 120.

\bibitem[{Jordan et~al.(1999)Jordan, Ghahramani, Jaakkola, and
  Saul}]{jordan1999introduction}
Jordan, M., Ghahramani, Z., Jaakkola, T., and Saul, L. (1999).
\newblock \enquote{An introduction to variational methods for graphical
  models.}
\newblock {\em Machine learning\/}, 37(2): 183--233.

\bibitem[{Kim et~al.(1998)Kim, Shephard, and Chib}]{kim:98}
Kim, S., Shephard, N., and Chib, S. (1998).
\newblock \enquote{Stochastic Volatility: Likelihood Inference and Comparison
  with {ARCH} Models.}
\newblock {\em The Review of Economic Studies\/}, 65(3): pp. 361--393.

\bibitem[{Knowles and Minka(2011)}]{ncvmp}
Knowles, D.~A. and Minka, T.~P. (2011).
\newblock \enquote{Non-conjugate Variational Message Passing for Multinomial
  and Binary Regression.}
\newblock In {\em Advances in Neural Information Processing Systems (NIPS)\/},
  25.

\bibitem[{Liesenfeld and Richard(2008)}]{LiesenfeldEIS}
Liesenfeld, R. and Richard, J.-F. (2008).
\newblock \enquote{Improving {MCMC}, using efficient importance sampling.}
\newblock {\em Computational Statistics and Data Analysis\/}, 53(2): 272 --
  288.

\bibitem[{Lovell(2008)}]{lovell2008simple}
Lovell, M. (2008).
\newblock \enquote{A Simple Proof of the FWL Theorem.}
\newblock {\em The Journal of Economic Education\/}, 39(1): 88--91.

\bibitem[{Minka(2005)}]{epdivergence}
Minka, T. (2005).
\newblock \enquote{Divergence measures and message passing.}
\newblock Technical Report MSR-TR-2005-173, Microsoft Research.

\bibitem[{Minka(2001)}]{MinkaThesis}
Minka, T.~P. (2001).
\newblock \enquote{A family of algorithms for approximate {B}ayesian
  inference.}
\newblock Ph.D. thesis, MIT.

\bibitem[{Minka et~al.(2010)Minka, Winn, Guiver, and Knowles}]{InferNET10}
Minka, T.~P., Winn, J.~M., Guiver, J.~P., and Knowles, D.~A. (2010).
\newblock \enquote{Infer.{NET} 2.4.}

\bibitem[{Nemirovski et~al.(2009)Nemirovski, Juditsky, Lan, and
  Shapiro}]{robustsa}
Nemirovski, A., Juditsky, A., Lan, G., and Shapiro, A. (2009).
\newblock \enquote{Robust Stochastic Approximation Approach to Stochastic
  Programming.}
\newblock {\em SIAM Journal on Optimization\/}, 19(4): 1574--1609.

\bibitem[{Nickisch and Rasmussen(2008)}]{Nickisch2008}
Nickisch, H. and Rasmussen, C.~E. (2008).
\newblock \enquote{Approximations for Binary {G}aussian Process
  Classification.}
\newblock {\em Journal of Machine Learning Research\/}, 9: 2035--2078.

\bibitem[{Nott et~al.(2012)Nott, Tan, Villani, and Kohn}]{nottstochastic}
Nott, D., Tan, S., Villani, M., and Kohn, R. (2012).
\newblock \enquote{Regression density estimation with variational methods and
  stochastic approximation.}
\newblock {\em Journal of Computational and Graphical Statistics\/}, 21(3):
  797--820.

\bibitem[{Opper and Archambeau(2009)}]{archambeau}
Opper, M. and Archambeau, C. (2009).
\newblock \enquote{The Variational {G}aussian Approximation Revisited.}
\newblock {\em Neural Computation\/}, 21(3): 786--792.

\bibitem[{Ormerod and Wand(2010)}]{ormerod:2010}
Ormerod, J.~T. and Wand, M.~P. (2010).
\newblock \enquote{Explaining Variational Approximations.}
\newblock {\em The American Statistician\/}, 64(2): 140--153.

\bibitem[{Paisley et~al.(2012)Paisley, Blei, and Jordan}]{paisleyvb}
Paisley, J., Blei, D., and Jordan, M. (2012).
\newblock \enquote{Variational {B}ayesian Inference with Stochastic Search.}
\newblock In {\em International Conference on Machine Learning 2012\/}.

\bibitem[{Porteous et~al.(2008)Porteous, Newman, Ihler, Asuncion, Smyth, and
  Welling}]{porteous2008fast}
Porteous, I., Newman, D., Ihler, A., Asuncion, A., Smyth, P., and Welling, M.
  (2008).
\newblock \enquote{Fast collapsed {G}ibbs sampling for latent {D}irichlet
  allocation.}
\newblock In {\em Proceedings of the 14th ACM SIGKDD International Conference
  on Knowledge Discovery and Data Mining\/}, 569--577.

\bibitem[{Richard and Zhang(2007)}]{RichardEIS}
Richard, J.-F. and Zhang, W. (2007).
\newblock \enquote{Efficient high-dimensional importance sampling.}
\newblock {\em Journal of Econometrics\/}, 141(2): 1385 -- 1411.

\bibitem[{Robbins and Monro(1951)}]{robbinsmonro}
Robbins, H. and Monro, S. (1951).
\newblock \enquote{A Stochastic Approximation Method.}
\newblock {\em The Annals of Mathematical Statistics\/}, 22(3): 400--407.

\bibitem[{Saul and Jordan(1996)}]{saul1996exploiting}
Saul, L. and Jordan, M. (1996).
\newblock \enquote{Exploiting tractable substructures in intractable networks.}
\newblock {\em Advances in Neural Information Processing Systems\/}, 486--492.

\bibitem[{Stern et~al.(2009)Stern, Herbrich, and Graepel}]{stern09}
Stern, D.~H., Herbrich, R., and Graepel, T. (2009).
\newblock \enquote{Matchbox: large scale online {B}ayesian recommendations.}
\newblock In {\em Proceedings of the 18th International Conference on World
  Wide Web\/}, 111--120.

\bibitem[{Storkey(2000)}]{Storkey00dynamictrees}
Storkey, A.~J. (2000).
\newblock \enquote{Dynamic Trees: A Structured Variational Method Giving
  Efficient Propagation Rules.}
\newblock In {\em Conference on Uncertainty in Artificial Intelligence
  (UAI)\/}.

\bibitem[{Teh et~al.(2006)Teh, Newman, and Welling}]{teh2007collapsed}
Teh, Y., Newman, D., and Welling, M. (2006).
\newblock \enquote{A collapsed variational {B}ayesian inference algorithm for
  latent {D}irichlet allocation.}
\newblock {\em Advances in Neural Information Processing Systems\/}, 19:
  1353--1360.

\bibitem[{Turner et~al.(2008)Turner, Berkes, and Sahani}]{Turner}
Turner, R.~E., Berkes, P., and Sahani, M. (2008).
\newblock \enquote{Two problems with variational expectation maximisation for
  time-series models.}
\newblock In {\em Proceedings of the Workshop on Inference and Estimation in
  Probabilistic Time-Series Models\/}, 107--115.

\bibitem[{Wainwright and Jordan(2008)}]{wainwright2008graphical}
Wainwright, M.~J. and Jordan, M.~I. (2008).
\newblock \enquote{Graphical models, exponential families, and variational
  inference.}
\newblock {\em Foundations and Trends{\textregistered} in Machine Learning\/},
  1(1-2): 1--305.

\bibitem[{Winn and Bishop(2006)}]{Winn2006}
Winn, J. and Bishop, C.~M. (2006).
\newblock \enquote{Variational message passing.}
\newblock {\em Journal of Machine Learning Research\/}, 6(1): 661.

\end{thebibliography}

\end{document}